\documentclass[pra,twocolumn,showpacs,amsmath,amssymb]{revtex4}
\usepackage{epsfig}
\usepackage{color}
\usepackage{graphicx}
\usepackage{bm}
\usepackage{mathtools}
\usepackage{gensymb}
\usepackage{xcolor}

\usepackage[colorlinks,bookmarks=false,citecolor=blue,linkcolor=red,urlcolor=blue]{hyperref}

\begin{document}

 \title{Satellite-based Quantum Information Networks: Use cases, Architecture, and Roadmap}

\author{Laurent de Forges de Parny$^1$, 
Olivier Alibart$^2$,
Julien Debaud$^{3}$,
Sacha Gressani$^{3}$, 
Alek Lagarrigue$^{2,1,4}$, 
Anthony Martin$^2$, 
Alexandre Metrat$^{3}$, 
Matteo Schiavon$^5$, 
Tess Troisi$^{2,1}$, 
Eleni Diamanti$^5$, 
Patrick G\'elard$^{4}$,
Erik Kerstel$^{3,6}$,  
S\'ebastien Tanzilli$^2$ and 
Mathias Van Den Bossche$^1$}

\affiliation{$^1$ Thales Alenia Space, 26, avenue J-F Champollion, 31037 Toulouse, France}
\affiliation{$^2$ Universit\'e C\^ote d'Azur, CNRS, Institut de Physique de Nice (INPHYNI), UMR 7010, Parc Valrose, 06108 Nice Cedex 2, France}
\affiliation{$^3$ Centre Spatial Universitaire de Grenoble, 120 Rue de la Piscine, 38400 Saint-Martin-d'H\`eres, France}
\affiliation{$^4$ Centre National d'Etudes Spatiales, 18 Av. Edouard Belin, 31400 Toulouse, France}
\affiliation{$^5$ Sorbonne Universit\'e, CNRS, LIP6, 4 Place Jussieu, F-75005 Paris, France}
\affiliation{$^6$ Universit\'e Grenoble Alpes, CNRS, Laboratoire Interdisciplinaire de Physique, 140 Rue de la Physique, 38400 Saint-Martin-d'H\`eres, France}

\date{\today}

\begin{abstract}
\textbf{ABSTRACT}. 
Quantum Information Networks (QINs) attract increasing interest, as they enable connecting quantum devices over long distances, thus greatly enhancing their intrinsic computing, sensing, and security capabilities. The core mechanism of a QIN is quantum state teleportation, consuming quantum entanglement, which can be seen in this context as a new kind of network resource. Here we identify use cases per activity sector, including key performance targets, as a reference for the network requirements. We then define a high-level architecture of a generic QIN, before focusing on the architecture of the Space segment, with the aim of identifying the main design drivers and critical elements. A survey of the state-of-the-art of these critical elements is presented, as are issues related to standardisation. Finally, we explain our roadmap to developing the first QINs and detail the already concluded first step, the design and numerical simulation of a Space-to-ground entanglement distribution demonstrator.
\end{abstract}

\maketitle

\section{Introduction}

This era witnesses the arrival of new technologies based on the manipulation and control of a few quantum objects, sometimes referred to as the 'second quantum revolution'. Like the 'first quantum revolution' that introduced electronics and photonics half a century ago, this second revolution is likely to have a profound impact on our society. Manipulating quantum objects means modifying the quantum state of these objects, and as these states convey quantum information, this in turn means encoding information at the lowest possible physical level. In the quantum realm,  properties like state superposition (i.e., contextual inderterminism)\cite{KochenSpecker,grangierAuffeves,Peres}, quantum non-local correlations (entanglement) \cite{BellInequalities} and no-cloning of an arbitrary unknown quantum state, enter into play. The technologies that exploit this ability to manipulate the state of individual quantum objects can be classified into three different categories, computing (annealers, noisy processors), sensing (cold atom accelerometers, optical clocks, magnetometers) and communications (quantum information networks, cryptography). The latter category is the object of this paper, and may lead to a new kind of communication network aiming to transport quantum information \cite{Nielsen_Chuang_2011, Watrous_2018, Kimble_2008, diamanti_2016, Wehner_2018, Gisin_2002, Pirandola_2020, Giovannetti_2011}.

The exploitation of entanglement to build general-purpose quantum-communication networks that will connect quantum devices in a network, such as quantum computers, quantum sensors or quantum cryptographic-key devices is under active study \cite{Kimble_2008, Wehner_2018, Gisin_2002, Pirandola_2020, Dahlberg_2019}. Since these networks will transport quantum information, it is convenient to call them simply quantum-information networks (QINs). Quantum teleportation is the core mechanism of these QINs \cite{Ottaviani_2015, Razavi_2018}. It requires establishing long-distance entanglement relationships between remote end-users. QINs are the basic technical objects required to create these relationships: they are entanglement establishment networks. The resulting communication between any pair of end users is intrinsically secured by entanglement monogamy \cite{Osborne_2006, Sheng2021}. Thanks to the many features and capabilities they will enable (including remote quantum computing, secure multiparty computation, extended quantum sensors, anonymous voting, quantum money, cryptographic keys, and others yet to be invented), such QINs are expected to result in a paradigm shift in communications, offering innovative services.

Without loss of generality, we consider that the role of a QIN is to enable teleporting states of qubits between two distant end-users at their respective network access points. The QIN shall provide entangled quantum systems (e.g., entangled photon pairs) to these access points, whose state is consumed by the teleportation operation. In this sense, entanglement is the basic network resource of the QIN, as the electromagnetic spectrum is the basic resource of classical networks. So the functionalities of a QIN aim at producing, transmitting  and exploiting entanglement. This also requires auxiliary functions such as, e.g., high performance synchronisation (time stamping)\cite{dauria_npj}. 

At this point, one has to consider the distance between the QIN end users as it will strongly impact communication performance. Here, we consider that the distance between users could be on a global scale. Quantum signals are intrinsically very weak and the no-cloning theorem prevents us from amplifying them to compensate for the losses, contrary to the signal in classical networks. The system  must therefore tolerate losses. Practically, in a fibre the transmission losses increase exponentially with distance, while they follow a square-law relationship in free space. As a result practical distance limitations of direct, real-life deployed fibre links are of some hundred kilometres \cite{Martin_2021}, while free space links enable direct links of thousands of kilometres \cite{Micius_entanglement_2017,Micius_teleportation_2017}. This range can be extended by introducing entanglement switches, i.e., relay nodes between such elementary links that swap entanglement, thus increasing the complexity of the overall network. To contain the complexity with as few switches as possible in the network, as well as to bridge over natural barriers (e.g. seas) or reach fibreless remote places, it is convenient to consider long elementary links provided by free space satellite nodes \cite{Picchi_2020, Chiti_2021, Chiti_2022, Wallnofer_2021}.
So for QINs as for classical communications, satellites provide a solution to reach service ubiquity \cite{Sidhu_2021}.

Although satellites allow QIN ubiquity, free space optical links use visible and near infra-red wavelengths that are strongly affected by atmospheric phenomena (clouds, aerosols, and atmosphere turbulence). Therefore, the provision of entanglement resource by satellites shall be decoupled from the user requests and subsequent consumption. One solution is to use the satellite as a prior provider of entanglement resource that will be stored on the ground to be used when needed. Once the entangled photons are stored in ground nodes, the nodes will be in a position to swap entanglement to the end users upon request, such that the end users can use this entanglement to eventually perform quantum state teleportation between them. This storage requires long-term and easy-to-address quantum memories. Although much progress has been made in the field of quantum memories, operational devices with the required characteristics remain to be matured to the level needed for operations \cite{Brennen_2016, Heshami_2016}. 

In the present paper, we focus on such a satellite-based QIN, arguably the most ambitious goal of quantum communications, while  remaining close enough to the current state-of-the-art, such that is can reasonably be expected to play a role in the telecommunication industry in the mid term. The QIN of interest in this paper has the purpose of enabling operational quantum communications services on a global scale to interconnect quantum computers, quantum sensors, and secured communication devices.


\section{Results}
\label{sec_results}


\subsection{Use cases of a Quantum Information Network}
\label{sec_usecases}

\subsubsection{Identified use cases per activity sector}
\label{sec_Identified_use_cases}

The activity sectors that could benefit from the services of a QIN are at least the following ones: 
industry (automotive, chemical, etc), critical infrastructures, finance, administration, operational science, and fundamental science.
In the following paragraphs, we explain how each sector could benefit from QIN services.
\\
\\
In the industrial field, a number of players have already expressed their interest in using quantum delegated or blind computing capabilities. Their objective is generally to take advantage of the capabilities of these computers to solve optimisation problems relative to their business sector, or to represent physical systems of interest more accurately. In this respect we mention: The  mechanical industries, for example the automotive/aeronautical sector, for the optimisation of finite element models; the pharmaceutical industry, to simulate larger and more complex molecules; the materials industry, in order to predict the properties of materials, which are ultimately related to quantum physics. In addition to these industrial sectors, the High-Performance Computing (HPC) capacity providers themselves, who will be able to offer quantum HPC services to industry to meet all of the above mentioned needs through adapted interfaces. Beyond blind computing, HPC actors will be interested in the capability to distribute computing tasks over several machines in order to leverage more processing resources. In addition to quantum computing, industries will certainly be interested in the possibility of establishing encryption keys to protect trade secrets.
\\
\\
Under the term critical infrastructure, we group together the various large systems deployed on the scale of countries or continents  that provide the structural means for the functioning of a society, e.g., power generation and distribution systems (electric, gas, etc.), water distribution networks, public transportation systems,
health data centres, telecommunication networks, and specific systems for managing critical infrastructures or monitoring the environment to anticipate major crises (such as flood control in The Netherlands, seismograph networks in Japan, pollution sensors, etc.).
\\
Within these critical infrastructures, we have identified two types of use. On the one hand, many critical infrastructures need precise timing systems for their operations, and the required accuracy increases with the complexity of these infrastructures. This is either to coordinate operations at a very high frequency, or to be able to date events in order to identify incipient cascades of events that could lead to a system blockage. For instance, the growth in throughput of communication networks requires ever more precise time control of the sequence of execution of tasks at their nodes. Another example is the timing of abnormal voltage variations in an electrical distribution network, which makes it possible to cut off certain links if they can be identified, before the anomaly propagates until the network collapses. The increasing need for accuracy will eventually lead to accuracies that only quantum sensors will be able to satisfy, and quantum sensor networks will emerge naturally. On the other hand, these infrastructures all have a command/control service for their distributed elements, and the intervention of a malicious actor in these services can block the infrastructure, which is a way to jeopardise the society that benefits from this infrastructure. Protecting this service relies, among others, on securing communications (at least authentication), which a QIN will also be able to do. Major telecom operators seriously consider securing their networks in this manner.
\\
\\
Banks and the financial sector are critical infrastructures of a particular kind, in the sense that if they share the need for precise timing of transactions and the need for securing their communications, and their general evolution leads them to use ever more computing resources. The key driver for this evolution is the movement of tokenisation of finance assets, whose most visible effect is the appearance, and then the generalisation of cryptocurrencies -- not the already existing private cryptocurrencies, but the ones that central banks are preparing, and which are planned to become the dominant means of exchange in our economies. 
\\
As part of such an infrastructure, quantum information networks have two roles to play. i) Enable distributed computing, as the authentication of transactions within a blockchain will most certainly be accelerated by the use of Grover-type quantum algorithms \cite{Grover1996}, and ii) The global security of processing and storage assets. Indeed, in such systems, the risk of a security breach that allows a malicious actor to steal or destroy value is no longer evaluated at the order of hundreds of millions of euros, as is the case with current attacks on Bitcoin. The level of risk of a security breach on central cryptocurrencies is on the scale of a country's economy or that of the entire planet, 3 to 6 orders of magnitude higher than the current outstandings with the existing cryptocurrencies.  Such stakes are likely to attract players on a whole other level, with far greater means of attack than current cryptocurrency offenders. The direction that central banks are taking is to use only information-theoretically secured means, among them quantum keys.
\\
\\
Concerning administration, we group under this term the means at the disposal of the political management organisations of our societies. These administrations are in charge of very vast domains (elections, diplomacy, justice, etc.) that are likely to benefit from quantum telecommunication services. We can cite as examples,  electoral processes, which could benefit from perfectly anonymous and perfectly reliable voting systems based on quantum protocols \cite{LiZeng, XueZhang} - at least in democratic societies, and authentication processes of digital documents for the validity of contracts, mandates and diplomatic communications (e.g. comparison of documents without exchange, encryption). These are example of applications of QINs that fall within the broader domain of secure and/or anonymous communications and that will go well beyond what is currently possible.  
\\
\\
For operational science, we distinguish the case of operational science from that of fundamental research. While the latter is concerned with the fundamentals of our scientific knowledge, the former has as goal the implementation of systems that rely on advanced knowledge to provide information that can be used in the very short term by societies (e.g., meteorology, geodesy, ...). For operational science, one can envision a large number of use cases for a QIN: services that rely on heavy computational activity (resolution of optimisation problems in geodesy, positioning or massive parallel computation in weather forecast) will certainly be interested in the use of distributed and probably also blind computation; Entities providing time reference services will certainly use QINs to improve performance in terms of stability and accuracy of their services though a more precise synchronisation of reference clock time, as well as of distributed time. Geodetic services will also be able to use clocks synchronised by such means to implement chronometric geodesy means; large baseline interferometric systems of electromagnetic or gravity quantum sensors to further increase their sensitivity for the benefit of remote sensing and environmental monitoring.   
\\
\\
Finally, the field of fundamental science will probably make use of all possibilities offered by QINs, since this activity is so demanding in terms of processing resources, measurement and detection, and means of communication. On the other hand, it is difficult to be more precise in the case of such an evolving, innovation-rich, and diverse domain, in particular on the time horizon that concerns the operational implementation of a QIN. It should be noted that the QIN itself is likely to become a means of scientific observation, like a large distributed instrument, because of its great sensitivity to very subtle phenomena \cite{Belenchia_2022}.

\subsubsection{Key performance indicators of a QIN}
\label{sec_Key_performance_parameters}

This section identifies the key performance parameters (KPIs) that characterise the functioning of the QIN. These parameters will be objects of contractual agreement between the QIN service suppliers and their clients. They will be derived and allocated as requirements for the subsystems of the network.

i) Latency, measured in seconds, is the time it takes to prepare the end users so that they can transmit their quantum information.
The transmission time itself depends on the performance of the end-user equipment and is not a characteristic of the network. So latency depends mainly on the end-to-end routing time of the entanglement through the network, the properties of the auxiliary channel and the reactivity of the network control \cite{Picchi_2020, Chiti_2021, Chiti_2022}. It does not include the time to create entanglement on elementary network links (via satellite or fibre) because this is done (especially for satellites) prior to the user request, so that the satellite does not have a real-time role in the system. Indeed, the need for availability of the service requires that one does not depend on the geometrical or meteorological visibility of the satellite. As such, the satellite is used to create supplies of entanglement resource, thanks to sufficiently stable memories on the ground, this resource being consumed later during quantum communications. 

ii) Entangled state fidelity between two access points to the QIN is a quantity (usually expressed in \%) that describes how close the delivered end-to-end entangled state is to the intended, maximally entangled state \cite{Schumacher_2016}. It conveys the level of similarity (overlap) between transmitted and received states.

iii) Timing precision of the output interfaces of the service access points is 
the 2-$sigma$ precision of the time stamps in the system reference time scale required to discern two successive states, measured in seconds.

iv) Network Access Area defines the geographic coverage of the QIN for a given level of performance and  corresponding access point characteristics. 

v) Resource rate is the number of end-to-end entangled quantum systems available at the user access interfaces of the network, with the required fidelity and dating accuracy, over a given elementary duration, measured in qubit per second (qbps).

vi) Auxiliary channel capacity is the classical information throughput of the auxiliary channel that is needed to transmit protocol control information, measured in bits per second (bps).

vii) Quantum channel capacity is the maximum number of qubits that can be transmitted over a noisy quantum channel, measured in qubit per second. \cite{Lloyd_1997, Gyongyosi_2018}.

viii) Average (resp., overall) availability is the probability of receiving end-to-end entangled quantum systems available between two given (resp., any two) terminal nodes with the required rate, fidelity, timing and auxiliary capacity conditions.

\subsubsection{Reference performance requirements for a  QIN}
\label{sec_Key_performance_requirements}

\begin{table*}[t]
\caption{Estimated order of magnitude for the main key performance indicators (KPIs) of a QIN. The Network Access Area is not mentioned as it is up to the telecommunication operator to define it as a function of the targeted market.}
  \centering
	\begin{tabular*}{\linewidth}{@{\extracolsep{\fill}}llllllll}
                  \hline
		 \hline
		 Sector			   	         & Latency		& Fidelity 			& Dating              & Resource rate         	&  Auxiliary channel capacity    & Availability   \\
		     				       		&(s) 		         &  (\%)			&                          &  (kqbps)           	 &    (kbps)                   &  (\%)   \\
		 \hline
		 \hline
		Industry                             		& 1		        &  99.99 	                & $<$1ms             &  1  	&    100  	&   90 \\
		Critical Infrastructures		& 0.1		        & 99		                & $<$1 $\mu$s     & 0.01    &    1	&   99  \\
		Finance	                         		& 0.01		& 99		                & $<$0.1ms          &  0.1 	 &    10	&  99.999  \\
		Administrations	               		& 0.0001		& 99.99		       & $<$0.1 $\mu$s   & 1,000   &  100,000  &  99  \\
		Operational Science   		& 1			& 99.99		       & $<$ 1 $\mu$s     &  1  	&   100	 &   99 \\
		Fundamental Science		& 1			& 99.99		       & $<$ 1 $\mu$s     &   1	 &  100	  &  99  \\
		\hline
		\hline
	\end{tabular*}
	\label{tab1}
\end{table*}

We identify here, for each sector discussed at the beginning of the Results section,
situations of use case implementation, and we give order of magnitude estimates of the expectations that end users might have about the network performance. The figures should not be viewed as final, but rather as a first reference to be considered, both for the assessment of the user communities, as well as to set some objectives for the design of the first QINs. We consider here the QIN performance as seen by individual users through their access terminal, connected to a terminal node of the network.
\\
\\
Industry use cases are computation and communication. For computation a reference need can be the transfer of the state of a system of 1,000 spins-$\frac{1}{2}$ per computation job, for 10,000 computations jobs per day (to reach $\simeq$ 1\% error), with transfer latency of 1 s per computation. For communication a need could be the establishment of 1000 AES-512 keys per day per industrial facility.
Critical infrastructures use cases are sensing and communication. 
For sensing a reference need can be the establishment of one entanglement link per second between relevant pairs of clocks. For communication a need can be considered to establish 25 AES-512 keys per day per facility.
Finance has 3 use cases -- sensing (synchronisation), computation and communication. For sensing, in this case a reference need can be the establishment of one entanglement link per 100 seconds between two clocks. In communication one can consider the need to establish 10,000 blockchain authentications per hour. In computation, the need is to distribute the inputs of a Grover-like algorithm to mine a blockchain 10,000 times per hour.
Administration has two use cases related to confidential communications. One can consider a need to routinely establish 25 AES-512 keys per facility and day, and anonymously receive $10^7$ messages (e.g., ballots) of 10-bit length per hour (on, e.g., an election day). 
Operational science has 3 use cases  -- sensing, computation and communication.
For sensing, one can consider the need to establish one entanglement link per second to a remote clock. For communication, the need can be to establish 25 AES-512 keys per day per facility. For computation the need can be to transfer the state of a 1,000-spin-$\frac{1}{2}$ system per computation job, for 10,000 computation jobs per day (to reach $\simeq$ 1\% error), with transfer latency of 1 s per job.
Fundamental science also has 3 use cases -- sensing, computation and communication. For sensing, we consider the need to establish one entanglement link per second to another clock. For computation we take the need to transfer the state of a 1,000-spin $\frac{1}{2}$ systems per computation job, for 10,000 computation jobs per day (1\% error), with transfer latency of 1 s per job.
\\
\\
The estimated orders of magnitude for the QIN users' KPIs are summarised per sector in Table~\ref{tab1}. 
It has to be understood at this point that the authors invite potential users to open discussions on these values, either bilaterally, or through the Quantum Internet Research Group (QIRG) of the Internet Engineering Task Force (IETF) \cite{IETF}. 


\subsection{Architecture of a satellite-based Quantum Information Network}
\label{sec_architecture}

This section addresses general considerations on the organisation of a QIN including a satellite component. Also, specific aspects of the Space component will be detailed before outlining 
the state-of-the-art of the critical components and of the standardisation.
  
\subsubsection{High-level architecture of a QIN}
\label{sec_High_level_architecture}

\begin{figure}[t!]
\begin{center}
\includegraphics[width= 1 \columnwidth]{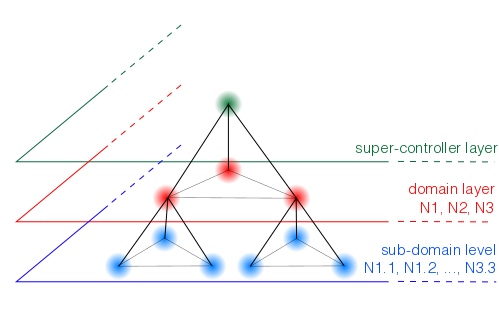}
\caption{Proposed subsidiary organisation of a QIN, reminiscent of software-defined networks. The figure shows the super-controller layer (green), the domain layer (red) and the sub-domain level (blue)} 
\label{Figure1}
\end{center}
\end{figure}

To deliver a QIN service that meets the user needs identified in the previous section, a QIN infrastructure needs to be deployed. 
\\
\noindent The boundary of the QIN infrastructure is taken to be the access point where entangled elements and auxiliary protocol data are delivered, thus including the delivery point, but excluding the entanglement-consuming devices that can be of many different kinds. This boundary sets the physical limit of the responsibility scope of the QIN service provider and manufacturer.
\\
\\ 
\noindent The QIN transfers quantum information by consuming the entanglement resource. The way this resource is produced and consumed follows three steps. First, entanglement is established along elementary links between neighbouring nodes that can be connected through fibre or free space links. Second, once a user request is received, end-to-end entanglement is established through entanglement swapping operations that consumes the resource on elementary links to create the end-to-end resource. Third, the end users consume the end-to-end resource to satisfy their specific need. 
Two conclusions have to be drawn here as far as architecture is concerned. First, there is a single end-to-end physical layer built upon entanglement. This is contrary to the case of classical networks where encoding in classical electromagnetic signals allows for encapsulation of information, leading to a complex concept of transport, as modeled in the classical Open System Interconnection (OSI) protocol stack. This OSI model does not apply here.
Second, quantum information does not hop from intermediate node to intermediate node, but directly from one end-node to the other end-node once end-to-end entanglement is established. In this sense, intermediate nodes route or switch entanglement and do not repeat (i.e., copy) any information. Contrary to their classical counterparts, they are switches, rather than repeaters.
\\
\\
Notwithstanding the need to create a protocol model that is different from the classical OSI stack (out of the scope of this paper), other classical network concepts remain relevant. As a matter of fact, one can still consider: 
i) A management plane for the QIN functions that are related to its (human) administration, 
ii)  A control plane to gather the automated functions that make its lower level functions work, and finally
iii) A resource plane that gathers the functions directly involved in the creation of the elementary-link and end-to-end entanglement resource ('resource plane' is preferred to 'data plane' because, as just mentioned, no data is transferred across the nodes of the network).   
\\
\\
\noindent The QIN needs to have a structuration that allows at the same time for flexibility of operations and for optimal management of the resource. These two characteristics are antagonistic in the sense that the former would naturally encourage a decentralised governance, while the latter favours a centralised governance. We propose here to follow a compromise that is a hierarchical organisation based on the subsidiarity principle illustrated in figure~\ref{Figure1}, reminiscent of software-defined networks. In this approach, the QIN is divided into parts that are autonomous domains (e.g., at the scale of a metropolis) and managed by a domain controller. When a domain receives a user request that it cannot manage on its own (e.g., connect with a user in a different domain), it resorts to a super-controller that manages a group of domains (i.e., the controllers and the interfaces of its subdomains).  The request is escalated until it reaches a controller common to both users, that manages the connection, defines the entanglement swapping route and flows it back down to the sub-controllers for implementation. Apart from potential domain border interfaces, only the lowest level domains own entanglement production and storage hardware.    
\\
More precisely, a domain is composed of end-nodes (users or interfaces to other domains), entanglement switch nodes with quantum memories, classical and quantum channels (with entanglement sources located in the middle of each link), subdomain managers and controllers. Each end-node is a user access point with a quantum access interface and potentially a quantum memory. These end-nodes are connected to one-another by classical channels (e.g. the classical Internet). 
\\
\\
An entanglement switch is a device located in a network node to transfer entanglement to one of several potential other nodes by entanglement swapping. Entanglement switches are connected to each other by both a quantum channel for entangled qubit exchanges and a classical channel for synchronisation, heralding \cite{DashiellVitullo} and unitary rectification (local operations and classical communications -- LOCC). The core of an entanglement switch is a Bell-state measurement device that measures jointly the state of one photon received from each of the two adjacent links to entangle the other photons of each link. We assume that each entanglement switch includes a quantum memory to store the resource and allow delayed swapping to a neighbouring node upon request. We also assume that a source of entangled photon pairs is located between each node of the QIN to create the quantum channel, and is thus part of the quantum channel that establishes elementary links. 
\\
\\
The subdomain controller ensures the operation and maintenance of the network and the resource management (transport, memory, computing capacity, physical resources mobilised, resource allocation, monitoring, billing, etc.). We assume that several sub-networks, although administrated by different operators, can be connected by both terrestrial (if geographically close) and Space nodes. The subdomain controller analyses the commands sent from the super network controller, measures and controls the quality of the service and differentiates the services (e.g., the nature of the service, topology, routing, notifications, throughput, synchronisation, requests, quantum bit error rate (QBER) correction, security, and user profile validation). 
\\
\\
Finally, the super-domain controller drives all the domain controllers. In our view, the super domain controller oversees and controls the entire QIN, including routing through the terrestrial and Space domains as needed, when an escalated request reaches its level \cite{Picchi_2020, Chiti_2021, Chiti_2022}. The super domain controller receives domain status information from the domains (active nodes, active links, resource levels, available nodes, resource production opportunities). The super domain controller will be in charge of establishing the routing plan for entanglement propagation paths across the subnetworks.
\\
\\
In this framework, we consider that there is no intrinsic difference between Space (that can be unified in a single domain or divided into several satellite domain operators) and terrestrial domains. Only the KPIs of the domains are likely to be different (e.g., larger coverage, but less capacity, in the case of a Space domain).

\subsubsection{Architecture of a Space-based QIN}
\label{sec_space_architecture}
 
\begin{figure*}
\begin{center}
\includegraphics[width=2 \columnwidth]{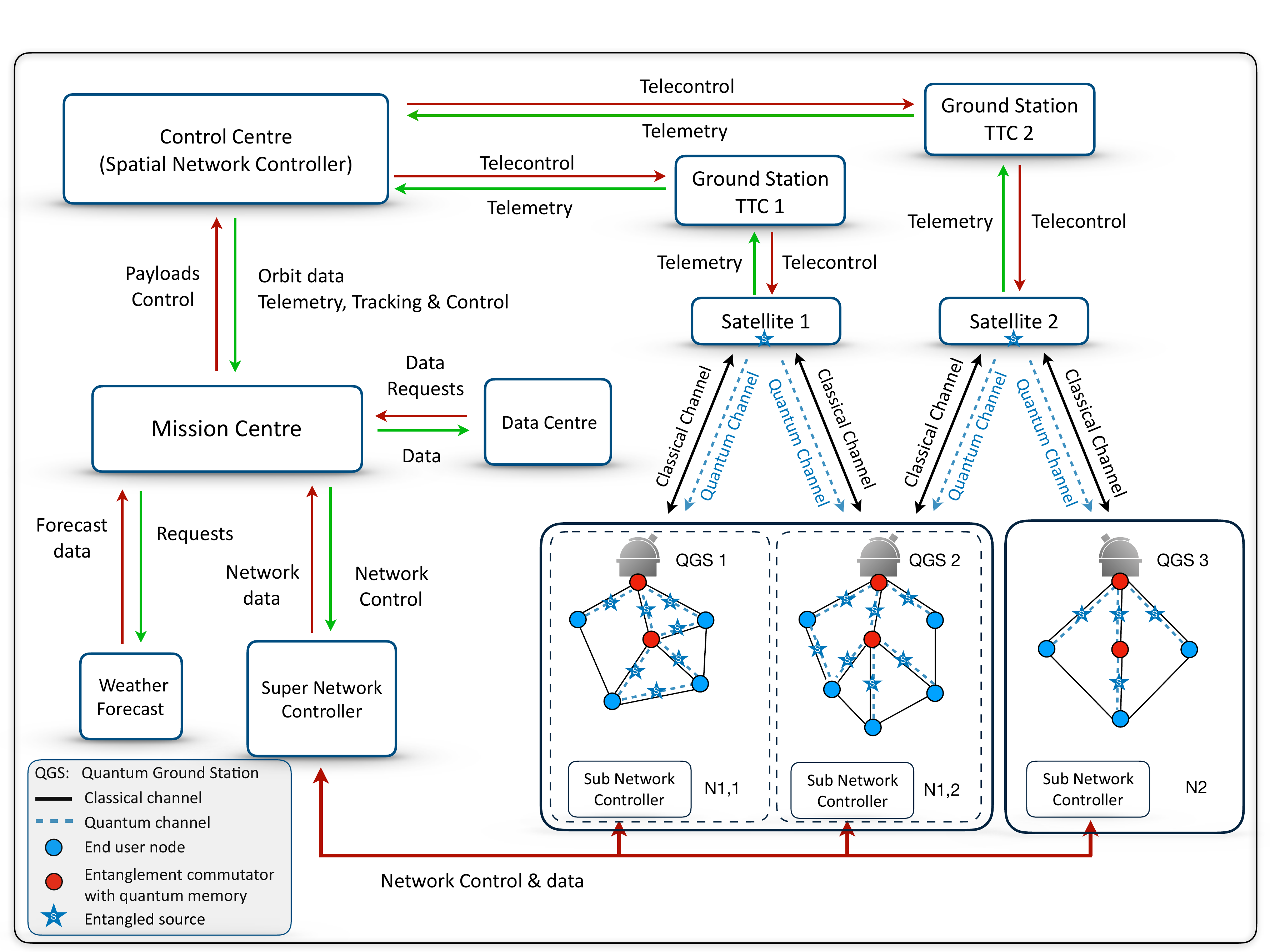}
\caption{Functionnal diagram of an integrated Quantum Information Network with Space segment. The sub- and super-network controllers manage the ground networks. The super-network controller communicates through the mission center with the control center in order to manage both the quantum and classical channels between the satellites and the Quantum Ground Stations (QGS).} 
\label{Figure2}
\end{center}
\end{figure*}

The high-level architecture of a Space QIN domain is composed of four main segments:
the Space segment, with assets in orbit;
the control segment, to command and control the Space segment;
the access segment, that enables the users to connect to the QIN, either directly or as part of a domain, and
the mission segment, that manages all the assets of the system, both on ground and in orbit, and also includes the domain controller for the Space domain.
A typical high-level architecture is depicted in figure~\ref{Figure2}.

The Space segment is composed of satellites. The satellites transmit pairs of entangled photons towards two receivers in the access segment, via two downlink quantum optical beams that create a single quantum channel between the two receivers. The satellite is thus a mid-point source. The Space segment also establishes classical communication channels (radio frequency or optical) between the satellite and the QGS receivers for related protocol data exchanges. Each satellite is composed of a payload and a platform. 

The payload comprises  the Entangled Photon Source (EPS), two optical terminals and a processor. The source of entangled photons produces entangled photon pairs in one of the four Bell states.  The source is equipped with a monitoring module for the on-board measurement of the transmitted quantum state fidelity, throughput, as well as the optical, thermal and mechanical status of the source. The on-board optical terminal is comprised of two telescopes for directing each photon of an entangled pair towards a scheduled station. The telescopes are mounted on remote control actuators for Pointing, Acquisition and Tracking (PAT) typically guided by beacon lasers. The payload processor analyses the commands, monitors the source status (power, optical, thermal, etc.), and potentially controls a laser master clock for ground-satellite synchronisation for time-stamping at sub-nanosecond scale. Furthermore, the payload processor optimises the encoding variable correction and ensures the self-calibration of the PAT device.

The platform includes the solar panels, the batteries, an on-board processor, a memory, a radio terminal for Telemetry, Tracking \& Control (TTC), sensors for satellite trajectory and attitude determination, actuators to modify the satellite attitude, thrusters, and a GNSS receiver. The platform manages the energy resources (solar panels, batteries), protects and stabilises the payload (heat, Space debris, radiations, vibrations), and maintains the satellite altitude and attitude. The platform sends telemetry data and receives commands from the control segment, and allows for the management of failures and de-orbiting at the end of the satellite's life.

The control segment, composed of the control centre and TTC stations, ensures the command and control of the satellites. It receives mission requests from the mission segment and converts them into accurate payload and platform commands. The remote commands are sent by the TTC stations connected to the control centre, and control the operations of the satellite payload and platform (power, photonics and electronics flows, equipment temperature, state of the entangled photon source, etc.). It also controls, and if necessary initiates correction of, the satellite orbit and attitude. The control centre receives and analyses satellite status information (telemetry), such as orbit status, trajectories, velocities, attitude, payload performance (throughput, fidelity of the entangled photons, temperature, etc.), processes ephemerides, and sends these data to the mission centre.

The access segment is composed of  QGSs that either directly serve end users, or connect to a local terrestrial QIN, or perform entanglement routing between two elementary links both fed by Space. Each QGS is composed of an optical terminal, a quantum receiver, an entanglement processor, an entanglement storage unit, a maintenance work-station, potentially a radio terminal, a secured ground network interfaced to the mission segment, and an entanglement switch at the interface with the local network or between elementary Space links that meet at the QGS, all hosted in a shelter below a cupola. Photons are collected though a telescope equipped with a PAT device, and their states are finally stored in a quantum memory until an entanglement swapping is performed to extend entanglement relationships and build end-to-end connections. The storage of entangled states allows the non-real-time and on-demand usage of the entanglement resource provided by the satellites. This key aspect ensures continuous operations of the ground network with available entanglement resources for end-users, independently of satellite availability and weather fluctuations that momentarily compromise optical communications between Space and ground. High accuracy timing systems, possibly in combination with cross-correlation of random photon arrival times at the two QGSs are used for time-stamping of the detected photons \cite{Scheidl_2009}.

Furthermore, the entanglement quality is normally increased through entanglement distillation \cite{distillation_2003, Entanglement_Purification_2021} by an entanglement processor. It first  discriminates against  spurious dark counts or background photons. More importantly still, it enables raising the fidelity of the entangled pairs, again at the expense of the number of available pairs, by correcting modifications of the components of the state accumulated on the path from the source to the receiver. Both error correction procedures require communication over a classical, but authenticated, channel between the two QGSs. 

Each QGS interfaces with the Space segment, the other QGSs, and the mission segment through classical channels (Internet/Ethernet) protected by a secured network front-end, as well as through quantum connections (entanglement switches).

The mission segment is composed of a mission centre, and is the master of the system.
In permanent conversation with the satellites and the access segment from which it receives status data, the mission centre is responsible for operations and builds the instructions that the control centre converts into platform and payload commands, tells the QGS which satellite to connect to, and shares information concerning the current and possible future status of entanglement resource with the super network controller. For these aspects, the mission centre takes into account weather forecasts and satellite ephemerides. The mission centre schedules the transmission of entangled pairs from a given satellite and instructs the corresponding QGS to prepare for reception. If swapping across several Space links is needed and possible, the mission segment serves as a domain controller to define the appropriate routing for this swapping. The mission centre is also responsible for processing mission monitoring data (performance level, quality-of-service policy, etc.) and for archiving operational data in a data centre, including TTC data received from the control centre, in order to allow for system troubleshooting. 

\subsubsection{Basic concept of operations for a QIN}
\label{sec_conops}

\begin{figure}[t!]
\begin{center}
\includegraphics[width=1 \columnwidth]{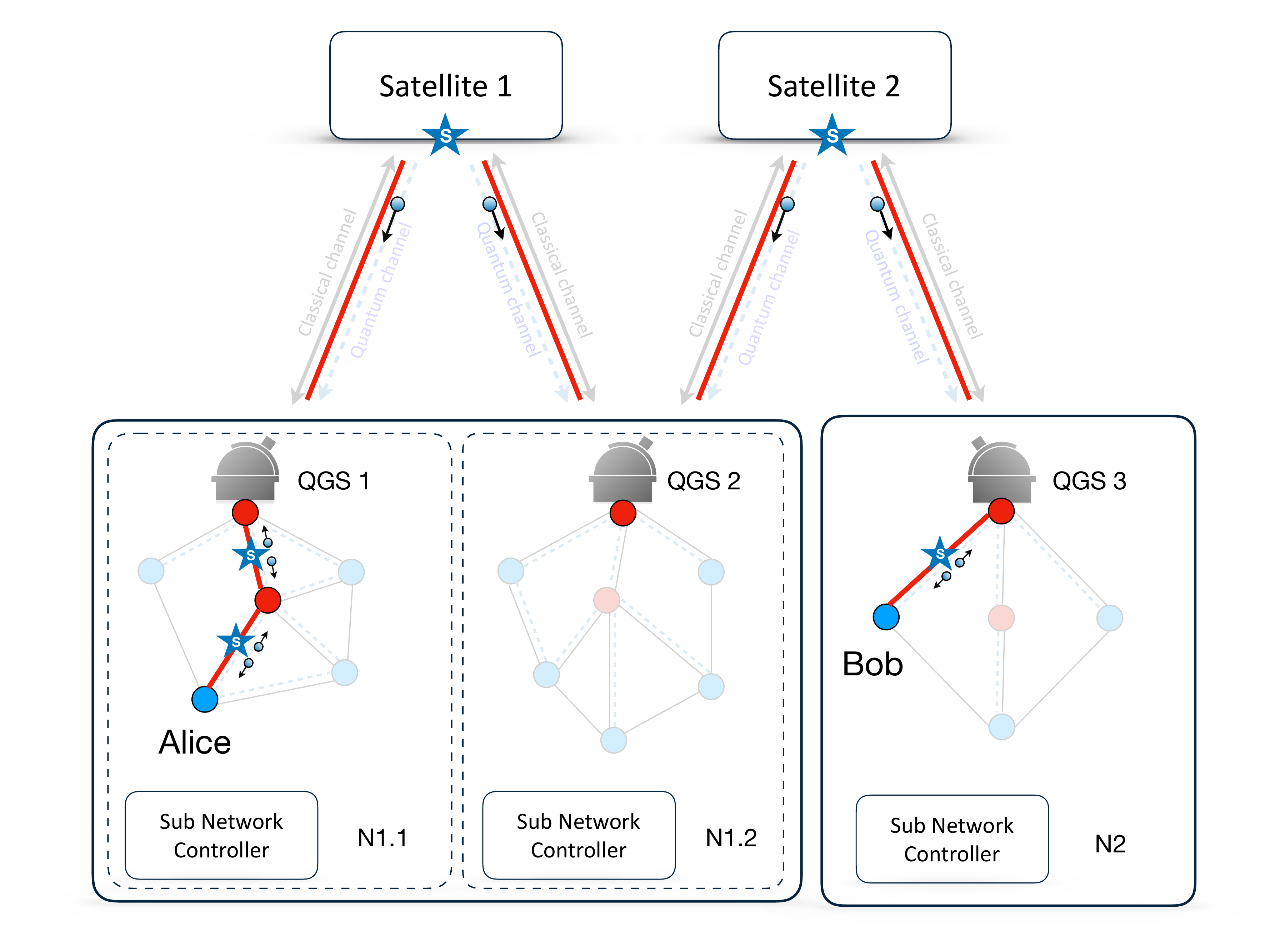}
\caption{Example of the establishment of a possible connection (red lines) between two QIN end-users, Alice and Bob (blue dots). The example involves multiple entanglement swappings between ground network (end-) nodes (blue and red dots) through entangled photon pair sources (blue stars) and two satellite-mediated connections between quantum optical ground station nodes in the networks.} 
\label{Figure3}
\end{center}
\end{figure}

We describe here a simple operational scenario that exemplifies the processing of a connection request between Alice and Bob, as shown in figure~\ref{Figure3}. 
Before any communication request by Alice or Bob, the network produces entanglement resource on elementary links, and stores it within quantum memories placed at each intermediate switching node. The satellites allow creating such elementary links over a long distance (typically $\gg$ 100 km). When the communication request is sent by two end-users, the appropriate level domain controller identifies the optimal path to connect these end-users, and organises the swapping of entanglement along this path. Red lines in figure~\ref{Figure3} show a possible path connecting Alice and Bob in a QIN. A Bell-state measurement allows the entanglement swapping to be performed at the switches and thus to weave an end-to-end entanglement link. Here, entanglement  follows the path indicated in red consisting of two satellites and four entanglement switches (red dots). The entanglement resource on each link will be consumed to build the end-to-end entanglement, and the end-to-end entanglement will be consumed at the moment of the communication between Alice and Bob, e.g., when Alice will teleport her qubit state to Bob.  We exemplify here domains (e.g., N1, N2  administrated by different operator entities), and subdomains (distant parts of the same operator domain, e.g., N1.1 and N1.2) of the QIN.

\subsubsection{Design drivers of Space-based QIN}
\label{sec_designDrivers}

The design choices of a Space-based QIN will be driven by a few structuring design drivers among which trade-offs have to be made. The main design drivers are:
i) the link direction: uplink or downlink.
ii) the satellite orbit altitude and inclination: Low Earth Orbit (LEO, i.e., 400-2000 km altitude), Medium Earth Orbit (MEO, i.e., 8000-25000 km altitude) or Geostationary Earth Orbit (GEO, i.e., 36000 km altitude).
If non-GEO, what inclination with respect to the equator plane?
iii) the aperture of the transmission and reception telescope primary mirrors: large on the ground and/or large on-board.
iv) the detector type: avalanche photodiode (APDs) or superconducting single-photon detectors (SNSPDs).
v) the quantum signal source wavelength: 810 nm or 1550 nm or higher.
These five design drivers are interconnected as they impact differently the link budget of the transmission, and will eventually determine the quantity of entangled resource available between the ground nodes.
\\
One more design driver is the quantum observable that encodes entanglement. We do not consider here continuous variable entanglement with squeezed states of light, because squeezing is quickly erased in the presence of channel losses, which are expected to be high in our case. In discrete variables, several options exist and none of them is excluded yet. For simplicity, we illustrate the concepts here with polarisation encoding, as it was used for the Micius \cite{Micius_entanglement_2017, Micius_entanglement_2020} and Socrates \cite{Socrates_2016, Socrates_2017, Takenaka_2017} satellite quantum communication demonstrations.
\\
\\
Link direction -- 
The choice of the link direction for the entanglement distribution depends on the possibility of embarking sources or detectors and quantum memories (QMs) on board satellites, as well as the difference in uplink vs. downlink losses (e.g., due to the so-called shower curtain effect, stringent pointing-ahead-, or timing requirements in the uplink). 
 QMs are currently being developed in research laboratories (see the subsection on "Critical subsystems of a QIN" here below), whereas SNSPDs, while providing high levels of performance ($\sim$90\% quantum efficiency), would require very significant efforts for their integration on board of a satellite, principally due to the necessity of cryocooling. 
 In the absence of Space-qualified QMs, synchronisation of the on-board Bell-state measurement involving two photons coming from two different ground stations is likely to be a practically insurmountable problem.
 Altogether, detection requires more resources and is easier to manage on the ground. 
Regarding propagation, atmospheric turbulence effects are easier to correct in the downlink configuration with adaptive optics on the ground using a downlink beacon laser for wavefront correction \cite{Pugh, Kerstel_2018}. For all these reasons, we chose a downlink quantum entanglement distribution scenario. 
\\
Satellite orbit -- The satellite altitude is a design driver with complex dependencies. Free space losses increase as the square of the distance between the satellite and the ground station. Conversely, coverage and satellite visibility duration increase with altitude. These in turn will call for fewer or more satellites in orbit to deliver a given level of service. This driver thus requires numerical analysis in order to find the satellite altitude and orbit that optimise the coverage and visibility time with acceptable resource rate. One solution will be given in the subsection "Towards a resource demonstration system".
\\
Size of the telescopes -- The link budget is controlled by the product of the on-board and on-ground telescope apertures. At this stage, one can only specify a value for the product of the two telescopes apertures and postpone the allocation of values to the Space and ground terminal apertures until economic factors are taken into account. As a matter of fact, large aperture terminals are obviously more costly, and large ones in Space might allow numerous cheap ones on the ground.
\\
Single photon detectors -- 
SNSPDs have better performance than APDs in a wide range of wavelengths of interest for quantum communications (from 200 nm to 2.2 $\mu$m)
 (see further down in the subsection "Critical subsystems of a QIN").
Nevertheless, they are cumbersome and expensive compared to APD detectors (for reasons of cryo-cooling, light-coupling, and wavefront correction requirements). It is therefore interesting to compare the system performance with both SNSPD and APD detectors. The comparison will be discussed in
the subsection "Towards a resource demonstration system".
\\
Quantum signal wavelength -- The choice of operating wavelength(s) of the quantum network is complex due to the large number of relevant phenomena that depend on it. The choice is also (often highly) constrained by the performance of the available products at each wavelength. A priori, the wavelength of the quantum signal used in the ground QIN fibre does not have to be the same as the one used for the Space-to-ground links. The requirements are far from being the same for these two components of the QIN. Only the Space component is subject to atmospheric absorption, scattering, turbulence, and strong background noise. Scattering strongly reduces atmosphere transmission at short wavelengths. At the same time, at very long wavelengths, an abundance of thermal photons prevents the use of single photon techniques. Previous studies have identified three near-infrared wavelengths that are compatible with detector maturity and the atmospheric transmission windows: 810 nm (or 850 nm), 1064 nm, and 1550 nm \cite{Micius_entanglement_2017, Micius_entanglement_2020, Socrates_2016, Socrates_2017, Takenaka_2017, ESA_2007, Vallone_2016, Liao_2017, Kerstel_2018}. It is important to note that the choice of the quantum signal source wavelength will determine the possibility of daytime operation. Indeed, the background noise from diffuse scattering, overwhelming during the day at 810 nm, does not allow operational quantum optical communications during daylight, whereas at 1550 nm this is considered to be possible \cite{Vallone_2016, Liao_2017, Avesani_2021}. 
Operation during day- as well as night-time will be a considerable advantage, particularly during summer where the duration of night time is strongly reduced at mid- and high-latitudes. Furthermore, the telecom C-band (1550 nm) is compatible with many optical devices and offers better single-mode optical fibre interface coupling efficiency than 810 nm. However, the beam divergence in free space, twice smaller for 810 nm than for 1550 nm, favours the shorter wavelength because it directly impacts the link budget. Furthermore, 810 nm enables direct light coupling to the surface of the Si-APD, thus avoiding the use of fibre coupling that introduces losses at the additional optical interfaces. Here we do not consider the alternative of 1064 nm for the following reasons:  it is not directly compatible with telecom technology, the wavelength is above the Si-APD cutoff, InGaAs (IGA) has a lower detection efficiency at 1064 nm than at 1550 nm (and a higher noise level than Si), and atmospheric transmission is lower at 1064 nm than at 1550 nm \cite{Carrasco_Casado_2020}. The comparison between the 810 nm and 1550 nm wavelength must be qualitatively investigated through numerical analysis. This will be discussed in 
the subsection "Towards a resource demonstration system".

\subsubsection{Critical subsystems of a QIN: State-of-the-art}
\label{sec_subsystems}

A number of subsystems will be indispensable for any implementation of a QIN. Here, we focus on the main quantum subsystems that are needed both in a ground QIN and on board the satellite payload: Entangled Photon Sources (EPSs), Single Photon Detectors (SPDs) and QMs.
\\
\\
Concerning entangled photon sources, the basic mechanism that generates the entangled photon pairs is Spontaneous Parametric Down Conversion (SPDC) that transforms a photon of the pump at wavelength $\lambda_0$ into two photons with wavelength $\lambda_1$ and $\lambda_2$, such that energy ($1/\lambda_0 = 1/\lambda_1+1/\lambda_2$) and momentum are conserved.  State-of-the art sources use mainly $\beta$-Barium Borate (BBO), Periodically Poled Lithium Niobate (PPLN) or Periodically Poled  Potassium Titanyl Phosphate (PPKTP) as non-linear crystals. The entangled photons can be either degenerate in energy, i.e.,  $\lambda_1=\lambda_2$, or not. 

The performance of the EPS is controlled among other things by fine-tuning the quasi-phase matching of the crystal to adjust the width and shape of the spectrum of the emitted photons. The entanglement can be encoded in time-bin, time-energy \cite{Franson_1989, Kwiat_1993}, or polarisation states \cite{Kwiat_1995, Kwiat_1999}. The optical power of the pump field within the crystal can be increased by placing the crystal in a  cavity that is resonant at the pump wavelength \cite{Oser_2020, Samara_2020}. Moreover, by adapting the cavity to the wavelength of the photons, it is then possible to reduce the spectral width of the emitted photons and make it compatible with the spectral acceptance of the QM, or to create a frequency comb to multiplex the beams at standard telecom wavelengths. To improve the efficiency of the non-linear interaction, another approach consists in confining the three fields (pump, signal, idler) over a long interaction length, which has been made possible by the development of waveguides in integrated photonics. This opens the possibility of high functional integration and reducing the complexity of the EPS \cite{Martin_2012, Ngah_Tanzilli_2015, Vergyris_2016}. 

As a result of the above evolution, two kinds of implementation can be considered for the source: compact bulk sources with free space optics in between optical components, and fibred sources where the beams are geometrically constrained by a propagation medium. 

For application in Space, the most advanced demonstrations use (bulk) EPSs on board satellites, as in the Micius and SpooQy satellites \cite{Bedington_2017}.

The Micius Space-borne EPS uses a continuous-wave laser diode at 405 nm with a line-width of 160 MHz to pump a PPKTP crystal inside a Sagnac-loop configuration \cite{Micius_entanglement_2017}, based on the architecture of \cite{Ramelow_2013}.  The crystal produces  down-converted polarisation-entangled photon pairs at 810 nm. With a pump power of 30 mW, an on-board rate of 5.9 million pairs per second with a fidelity of 0.907$\pm$0.007 was measured. In Micius, the entangled-photon pairs are then sent to the ground. 

In SpooQy \cite{Bai_2018, Bai_2020}, entangled-photon pairs are detected on board. As in Micius, SpooQy's Space-borne entangled-photon source uses a 405 nm continuous-wave pump laser with 160 MHz line-width. The polarisation-EPS is based on collinear, non-degenerate type-I SPDC with critically phase-matched non-linear crystals. The pump produces horizontally polarised photon pairs in two BBO crystals. Between the two BBO crystals, an achromatic half-wave plate induces a 90 deg rotation in the polarisation of the generated photons from BBO-1, while the pump polarisation remains unaffected. With 17 mW of pump power, the typical in-orbit detected pair rate was 2200 pairs/s with highest visibilities of 98\%.

Finally, the Fraunhofer IOF in Jena and IQOQI in Austria are developing a bulk source in the framework of the European Space Agency EPS-projects. A continuous laser at 405 nm pumps a PPKTP crystal in a Sagnac-loop configuration to generate entangled degenerate photon pairs at 810 nm with a performance of $6\times10^6$ pairs$\cdot$s$^{-1}$$\cdot$mW$^{-1}$ with a fidelity up to 98\% \cite{Steinlechner_2016, Beckert_2019}. This source recently passed many Space qualification tests (vibration, shock and thermal cycling). The group is now developing hybrid sources with entangled photons at 810 nm and 1550 nm and targets a $10^9$ pairs$\cdot$s$^{-1}$ rate. Pulsed sources are also under development \cite{Beckert_2021}.
\\
\\
Single photon detectors (SPDs) are mainly characterised by their detection efficiency or probability of detecting a photon, their dark count rate, recovery time after detection (or dead-time, limiting the maximum count rate), time jitter, cooling requirements (e.g., cryocooler or Peltier), purchase and operational costs, and their size, weight, and power consumption. Several technologies are used to detect light at the single photon level in Geiger mode, namely semiconductor-based APD or SNSPD.

There are two main families of semiconductor-based APDs: those made of silicon and those made of InGaAs (while APDs based on HgCdTe appear promising in several aspects, they are still at the development stage \cite{Rothman_2016, Rothman_2017, Rothman_2018}). Silicon detectors cover the visible range extended to the very near-infrared (NIR), i.e., 400-1000 nm. Generic silicon APDs have an optimised structure to detect NIR wavelengths (700-850 nm), with detection efficiencies up to 70\%. The purity of the silicon and the mastery of the manufacturing processes make it possible to produce detectors with a noise rate of the order of 30 to 300 dark counts per second, reaching 1 cps for the best detectors. For applications at telecommunications wavelengths, detectors based on the InGaAs compound semiconductor are more efficient than Si-APDs in the NIR domain ranging from 1000 to 1600 nm. However, their maximum efficiency is limited to 30\%. Since the breakdown voltage is lower than for silicon, it is easy to operate them in gated mode, which was the only possible mode of operation initially, due to a relatively high level of noise (typically of the order of 100 cps or higher). Nowadays, there are detectors on the market that can operate under gated conditions up to a frequency of 100 MHz, with pulse durations of 1 ns. Moreover, thanks to a lowering of the operating temperature to 163 K, noise rates of 1 count per second for an efficiency of 10\% have been obtained. The main drawback of these InGaAs detectors remains the dead-time, which is generally in the range from 10 to 100 $\mu$s, and thus imposes a maximum detection rate of $\sim$100 kHz.

The detection of single photons has been revolutionised by the emergence of SNSPDs based on superconducting niobium nanowires. The detection mechanism is based on the breakdown of superconductivity by the absorption of a photon that breaks Cooper pairs of electrons in the nanowire. The photons are then detected by measuring the variation in the resistance of the wire. This technology has been rapidly developing for several years now and a number of companies offer superconducting detectors based on different materials with slightly different specifications, such as for example an operating temperature between 0.8 and 3 Kelvin \cite{Caloz_2018}. SNSPDs have the advantage of operating over a wide wavelength range from the UV band up to the IR band (200 nm to 2.2 $\mu$m) and can achieve detection efficiencies up to 90\%. The noise is mainly due to black-body radiation of the room-temperature fibre that brings the photons to the SNSPD cryostat. To further reduce noise, it is possible to add narrow-bandwidth filters as close as possible to the detector so as to achieve noise rates lower than 1 count per second. The other interesting characteristics are the time jitter which is of the order of 20 ps, and the recovery time of the order of 100 ns to return to 90\% of nominal efficiency \cite{Caloz_2019, Autebert_2020}. Note that this recovery time is due to the electrical impedance of the wire, which slows the return of current to the superconductor. This impedance can be reduced by connecting several wires in parallel. In addition, this approach also allows producing detectors capable of resolving the number of photons per pulse. In general, SNSPD detectors offer a higher level of performance than semiconductor detectors, but this is accompanied by constraints related to the cooling of the active zone to cryogenic temperatures. This is done by means of closed-loop cryostats with helium recycling (cryocoolers) whose size (typically 0.1 m$^3$) and maintenance requirements have considerably reduced over the years.  

Finally, it should be mentioned that while single photon detectors are mature for terrestrial implementation, their operation on board a satellite remains a challenging task, particularly for SNSPDs, and for use in small satellites. Therefore, developments of Space compatible SPDs are oriented towards APDs without cryocooler, which will limit the single photon detection efficiency in Space and, consequently will promote the use of more efficient detectors on the ground, in a downlink configuration. 
\\
\\
Concerning QMs, although much progress has been made in recent years in the development of devices that will allow  storing entanglement over long durations, operational solutions remain elusive for the time being and only prototypes exist in research laboratories. However, concepts including on-board QMs are already being discussed for Space applications \cite{Gundogan_2021, Liorni_2021}.

The important characteristics of QMs are related to maximum storage duration and read/write ease:
\noindent i) Storage time, i.e., the maximum time between writing and reading into and out of the memory, which is limited principally by decoherence; in some simple understanding, it is sometimes identified as the time between photon absorption and re-emission;
\noindent ii) Fidelity of storage, which quantifies the similarity between the input written state and the output read state;
\noindent iii) Efficiency, i.e., the probability of success of a write/read operation, is linked to the probability that a photon is absorbed and re-emitted; 
\noindent iv) On-demand reading, i.e., the possibility to re-emit the photon at a precise, but not predetermined, time;
\noindent v) Multiplexing, i.e., the number of states that can be stored simultaneously into the memory;
\noindent vi) The interface wavelength, which should be compatible with the optimal wavelengths for photon transmission.
There are different protocols for memory implementations. Some memories are said to be ``emissive", without input state, while some others are ``absorptive", storing an input state that can be read later.  

The most successful example to date of emissive memories is based on the Duan, Lukin, Cirac et Zoller (DLCZ) protocol using cold atomic ensembles \cite{DLCZ}, which enables generating measurement-induced entanglement between two memories via a heralding event. A rudimentary version of a repeater segment was demonstrated using this approach \cite{Chou_2007}. In this seminal work, the retrieval efficiency was close to 50\%, but with a lifetime limited to about 20 microseconds. In a recent work \cite{Wang_2021}, the authors demonstrated the creation of atom-photon entanglement with exceedingly longer storage times. Their memory retrieval efficiency was 38\% for a storage time of 0.1 s and they managed to violate Bell inequalities after a storage time of 1 second.

``Absorptive" memories can rely on different physical platforms and protocols. The most used approaches are the Electro-magnetically Induced Transparency (EIT) protocol \cite{EIT_Fleischhauer}, mostly used with cold atomic ensembles, and the Atomic Frequency Comb (AFC) protocol \cite{AFC} used in doped crystals. Recent EIT implementations based on cold atomic ensembles led to record storage and retrieval efficiency, with storage of qubits and single photons with fidelities of 99\% and overall storage-and-retrieval efficiency close to 90\% \cite{EIT_Vernaz,Cao_2020}. In doped crystals, extension of AFC to on-demand storage is actively pursued \cite{Rakonjac_2021}, still with limited efficiency of about 6\% and a storage time up to 50 microseconds. Long storage times for an optical memory were demonstrated in different crystals, albeit not in the quantum regime. A coherence time up to 6 hours was demonstrated in an initial work \cite{Zhong_2015}, and storage up to 1 hour with 96\% fidelity using a $^{151}$Eu$^{3+}$:Y$_2$SiO$_5$ crystal with the AFC protocol was recently reported \cite{Ma_2021}. Pushing this capability to the quantum regime is still very challenging.  

The state-of-the-art for the entanglement distribution between distant QMs is given by two recent experiments \cite{LagoRivera_2021, Yu_2020}. Both experiments create atom-photon entanglement between two memories located at a long distance from each other, with telecom wavelength photons being sent to a third node where a Bell measurement is performed to entangle the two memories together. In the first experiment \cite{LagoRivera_2021},  QMs based on rare-earth doped crystals were used with an AFC protocol, in order to obtain a fidelity of 92\% for loss levels up to 6.5 dB per channel. The device was capable of storing photons for a time up to 25 $\mu$s and demonstrated the possibility to store up to 64 modes simultaneously in the same QM. The second experiment used two QMs based on cold atom traps and  the DLCZ protocol to create atom-photon entanglement \cite{Yu_2020}. The photons are then converted to the telecom wavelength and sent by optical fibres to a midway-station, where a Bell measurement transferred the entanglement to the two memories. It demonstrated the possibility of transferring entanglement over a total distance of 22 km using photons encoded in two modes of the electromagnetic field and a two-photon Bell measurement. A total distance of 50 km was reached for an encoding in Fock states and a single photon measurement.

In conclusion, while storage time, fidelity and addressability are improving, reading on demand and an output wavelength compatible with the communication network remain functionalities to be consolidated. 
As they are basic and essential ingredients of any large, functional QIN, QMs clearly require significant maturation.

\subsubsection{Status of standardisation} 
\label{sec_normalisation_standardisation}

Standardisation of the operating principles and the main interfaces is always a fundamental step towards commercial networks, so as to enable the development of a competitive ecosystem with solutions. Several international organisations have taken up the subject of quantum telecommunications, and standardisation is actively being studied. 
The main actors are:
The European Telecommunications Standard Institute (ETSI), of which the QKD Industry Specification Group (ISG) has been operating for a decade \cite{ETSI};
The International Standard Organisation (ISO);
The Internet Engineering Task Force (IETF), with the  Quantum  Internet  Research  Group (QIRG), active since 2017 \cite{IETF};
The International Telecommunication Union Telecommunication Sector (ITU-T) since 2018 \cite{ITU};
The Institute of Electrical and Electronics Engineers (IEEE) since 2019;
The European Committee for Electrotechnical Standardisation (CENELEC), very recently;
European Information Technologies Certification Institute (EITCI), with the Quantum Standards Group (QSG).

These groups do not all deal with the same subjects, although a certain level of overlap exists. Among the two main topics of quantum telecommunications, i.e., quantum secure networks (based on QKD) and QINs, it should be noted that the first one attracts in general more attention, due to the greater maturity of the concepts and the existence of a first generation of products on the market \cite{Alleaume_2014}. Among these groups, those that deal with the subject of QINs are the ITU-T, and especially the QIRG of the IETF.

The ITU-T work has identified a first and absolutely necessary task, which is to establish a general terminology for quantum communication networks (\cite{ITU} and references therein).  In doing so, it addresses both types of networks. The work of the QIRG is more specifically dedicated to QINs \cite{IETF}. For the past two years, the Internet Research Task Force (the sister organisation to the IETF) has hosted this QIRG that meets three times a year during IETF meetings, and interacts through online discussions in between. This group was formed  following the observation that the first devices that constitute the physical layer of a QIN are now becoming available. While these devices can still be significantly improved, the first connections 
they enabled suggest that it is time to think about the organisation of the network at a higher level. It is the objective of the QIRG to develop this organisation. The work within this group consists for the moment of preparing the documents that aim to refine the concepts necessary for the QIN.


\subsection{Roadmap to operational Space-based QINs}
\label{sec_roadmap}

It should be obvious from what is described in subsection "Architecture of a satellite-based QIN" that the way to operational Space-based Quantum Information Networks is still a long one \cite{Kaltenbaek_2021}. Yet all the elements required to reach this stage can now already be identified. This enables building a roadmap, of which the first step has already been executed (outlined in the following subsection "Towards a resource demonstration system"), so as to prepare the next steps towards
reaching the final goal of an operational QIN that delivers a service to end users as identified in subsection "Use cases of a Quantum Information Network".

\begin{figure*}
\begin{center}
 \hspace{-0. cm}{ \includegraphics[width=2.1 \columnwidth]{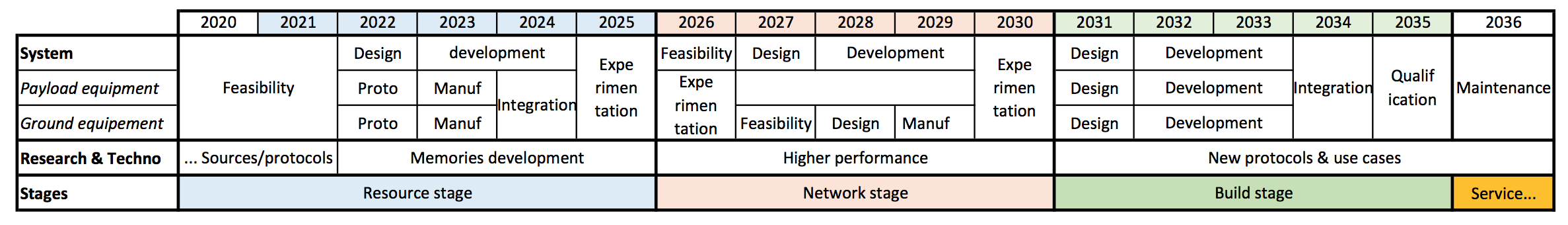}}
\caption{An outline of the roadmap that can be considered to reach the service opening of a first Space-based QIN. Some work of the resource stage has already been completed.}
\label{Figure4}
\end{center}
\end{figure*}

Our vision is that the roadmap towards a Space-based QIN as described in the subsection on the "Architecture of a satellite-based QIN"
has to be split in three stages: 1) demonstrate resource, 2) demonstrate networking, 3) build the first operational network, see figure~\ref{Figure4}.

The resource stage aims at demonstrating that the QIN resource can be distributed from Space to the ground.  Whereas the feasibility of the principle was demonstrated by the Micius team \cite{Micius_entanglement_2020}, the resource stage of this roadmap aims at demonstrating entanglement distribution from Space in a service oriented configuration. The objective is not only qualitative, but also quantitative: quantity, as well as the quality, of the entanglement resource will have to be demonstrated by a full scale experiment. 
This means deploying: a satellite that embarks a demonstration payload;  at least two prototype-QGS to collect the resource;
an experimental ground segment to assess the distribution performance though specific KPIs. 
Unfolding this process will allow finding solutions to all the technical challenges that need to be solved to enable a real-life implementation,  in order to demonstrate a global QIN, and to identify potential improvement tracks leading to an operational system (see the following subsection "Towards a resource demonstration system"). 
 
\noindent Critical during the resource stage is the maturation that has to be carried out in parallel on QMs, but also on efficient Bell-state measurement devices and on implementing entanglement heralding mechanisms. Together with the resource demonstration, this will allow moving to the next stage. 

The network stage builds on the resource demonstrator by upgrading the relevant elements of the demonstration system with the outcome of the matured technologies so as to allow assembling several QGS-QGS links through swapping operations into a first generation of QIN. This will enable building critical knowledge and know-how not only about the physical layer implementation, but also about the management of such a completely new kind of system. For instance, it can be expected that the optimal management of production and consumption of entanglement resource will lead to complex  problems that require novel solutions.   

The build-up stage will take all the experience accumulated in the first two stages and lead to the full design, manufacturing and deployment of a first operational infrastructure. This system will provide a first service to the users  identified in subsection "Use cases of a Quantum Information Network". This will probably require launching new satellites, upgrading QGSs already operational in quantum-key networks, such as 
the European Quantum Communication Infrastructure (EuroQCI), and of course deploying a full-fledged mission segment to manage the Space-based QIN mission.

Each of these stages shall involve the necessary skills to design the system and its components, to industrialise the components and to first assemble them into the demonstration system, to build its upgrade, and to finally complete the operational system. This means that from now on, a number of actors are expected to join the roadmap, from academic experts to large system integrators able to assume the high risk of building a new infrastructure, as well as a number of equipment-oriented SMEs that will supply the intermediate level building blocks of the QIN. 

Such a roadmap shall also identify the critical elements of the system, mainly based on their enabling character and their maturity, and plan an anticipated start of work on these critical elements. This is, for example, the case for QMs that have to be prepared for the network stage, while other actors work on the resource stage. At a lower level this is also the case for, e.g., the entangled photon source and for the quantum receiver equipment, where we have already anticipated some activities of the resource stage beyond the overall system design mentioned in the next subsection ("Towards a resource demonstration system").
   
As far as the timeline is concerned, 5 years can be considered a reasonable duration for the resource stage, if adequate  funding is available. Further stages are more difficult to plan at the time of writing, yet 5 years for each stage can be considered a reasonable estimate.

\subsection{Towards a resource demonstration system}
\label{sec_demonstrator}

We describe in this section the first results obtained after initiating the resource stage of the roadmap sketched out in the previous subsection ("Roadmap to operational Space-based QINs").
We first describe the system and then explain the trade-offs carried out among the various design drivers, and give first results on which to build the next steps of our roadmap.

\subsubsection{Demonstrator model and parameters} 
\label{sec_demonstrator_method}

\begin{figure}[t!]
\begin{center}
\includegraphics[width=1 \columnwidth]{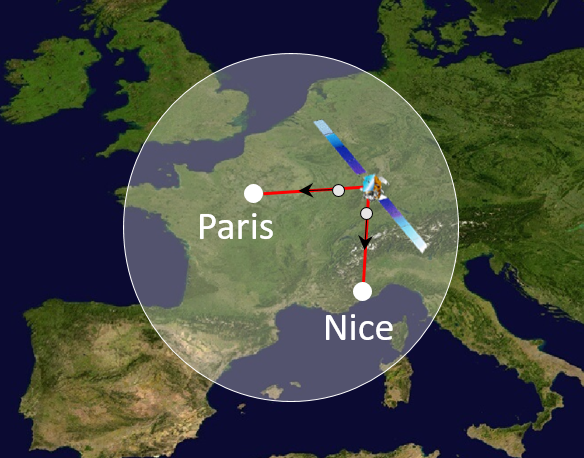}
\caption{Minimal system demonstrator. Space-to-ground entanglement distribution from a satellite towards two ground stations located in Paris and Nice, France.}
\label{Figure5}
\end{center}
\end{figure}

In this section we analyse the scenario of a minimal system to demonstrate entanglement distribution from Space. It is composed of one satellite and two QGSs located close to Paris and Nice and thus with a distance between the two QGSs of $\sim$680 km, see figure~\ref{Figure5}.

We consider here entanglement distribution in real time, i.e., without QM, in a downlink configuration. Entanglement is encoded  in the polarisation observable, although this might not be the final design choice.  We have performed numerical simulations of this configuration in order to address key trade-off decisions, discussed in subsection "Architecture of a satellite-based Quantum Information Network"
on the basis of a quantitative analysis. 
In particular, we compare the system performance, in terms of total link attenuation and entanglement distribution KPIs, as a function of the following system parameters: 
i) the satellite orbit (LEO or MEO, inclination), 
ii) the wavelength of the entangled photons (810 or 1550 nm), 
iii) the type of single photon detector (InGaAs-APD, Si-APD, or SNSPD).
We neglect the effects of clouds and aerosols in this study.

In order to perform this analysis, we have built a simulation tool that enables us to study the impact of different choices for the many parameters that define the system. This tool consists of three parts. The first is an orbital simulation package developed at the French Space Agency (CNES) known as Simu-CIC, running in the Scilab numerical computation environment \cite{CNES_SimuCIC}. It enables a mission scenario description including a detailed and realistic orbital propagation, taking into account, e.g., the  non-sphericity of Earth, the residual atmosphere drag on the satellite as a function of its weight, geometry, and orientation (attitude sequences), as well as the ground station locations. The starting time of the simulation and its time resolution are variable, but typically we start our simulations on July 1, 2021 with a time step of 10 seconds for simulations with a total duration of 12 months. One or more ground stations can be taken into account to enable the calculation of satellite elevation angle, altitude, and distance to the satellite. The orbital simulations and the various ephemeris files are stored on disk in the standard CNES "Centre d'Ing\'enerie Concurrente" (CIC) format  \cite{CIC_website}. These are subsequently read by a Python script that extracts the orbital sections during which the satellite is simultaneously visible by two selected QGSs. 
Here, visibility is defined by setting a minimum elevation angle ($\beta=30$ degrees), and by specifying whether the quantum channel between QGS and Spacecraft can operate during daytime and dawn/twilight, or only during nighttime. The duration of the dawn and twilight can be specified to include civil-, maritime-, and/or astronomical- dawn/twilight. 
During the night time section of the orbit, the detector background count rates are fixed to a predetermined value. In a future version we plan to make the background counts a function of the local time and geolocation.  A second Python program then calculates the atmospheric attenuation between the satellite and each QGS, followed by a calculation of the entanglement distribution KPIs. 

The atmospheric link simulation code is described in the Methods section. The input parameters are summarised in Table~\ref{tab2}.

\begin{table*}[t]
\caption{Simulation input parameters as discussed in subsection "Towards a resource demonstration system".}
  \centering
	\begin{tabular*}{\linewidth}{@{\extracolsep{\fill}}lll}
                  \hline
		 \hline
		 Parameter			   	         & Description		& Value    \\
		 \hline
		 \hline
		$\lambda$                                  & Quantum channel wavelength 		        &  810 nm / 1550 nm 	 \\
		$H$                            		& Satellite altitude        		                         &  LEO: 600 km / MEO: 8000 km \\
		$D_R$                            		& Ground receiver telescope diameter        		        &  80 cm @LEO / 100 cm @MEO \\
		$D_T$                            		& Onboard transmitter telescope diameter           		&  30 cm @LEO / 50 cm @MEO  \\
		$A_{atm, 0}$                              & Atmospheric  attenuation at zenith                 &  3 dB @810 nm, 2 dB @1550 nm	 \\
		$T_R$                            		& Receiver transmission factor    		        &  0.8	 \\
		$T_T$                            		& Transmitter transmission factor   		        &  0.8	 \\
$T_{optics}$ 			& Optical module transmission factor  &  0.2 @810 nm, 0.35 @1550 nm \\
		$L_p$                            		& Pointing  losses        		       		        &  0.3 @810 nm, 0.2 @1550 nm  	 \\
		$q$                            		& Basis reconciliation factor for BBM92       		       		        &  0.5 \\
		$f$                            		        & Bidirectional error correction efficiency     		       		        &  1.22\\
		$\tau$                            		&  Coincidence time window		       		        &  200 ps\\
		$\mu$                            		&  Average number of photon pairs per pulse	        &  0.02\\
		$D$                            		&  Detector dark count rate (same detectors in Paris and Nice)	        &  100 cps\\
		$B$                            		&  Background (stray light) count rate      		&  400 cps @810 nm, 100 cps @1550 nm	 \\
		$PDE$                            		&  Single photon detector efficiency      & 0.9(SNSPD)/0.68(Si-APD)/0.25(IGA-APD) \\
		$\eta_x$                      		 & Quantum channel efficiency		&  Eq.(\ref{eq:OverallChannelEfficiency})  \\
		$e_0$                      		         &  Error probability for dark- and background counts		& 0.5 \\
		$e_p$                      		         &  Error probability of photon arriving on wrong detector   & 0.01 \\
		$\Delta t$                      		 &  Time step   & 10 seconds \\
		$T_{final}$                      		 &  Total simulation duration   &  12 months \\
		\hline
		\hline
	\end{tabular*}

	\label{tab2}
\end{table*}

\subsubsection{Simulation results}
\label{sec_demonstrator_simulation_results}

Here, we present simulation results based on the modeling and parameters described in subsection "Towards a resource demonstration system".
Particularly, our results allow analyzing three trade-offs, namely the satellite orbit and altitude choice, the SPD choice, and the quantum signal source wavelength choice. 
The simulation package was first used to investigate the effects of the choice of satellite orbit, and notably its inclination. 
As expected, at the same height of 600 km, a LEO at an inclination of about 50 degrees increases the coverage of the locations of the selected QGSs at Nice (43 deg 42' N, 7 deg 15' E) and Paris (48 deg 61' N, 2 deg 21' E), compared to other orbital inclinations, and in particular, compared to a sun-synchronous orbit (SSO, inclination 97.8 deg). In order to investigate the effect of increasing the height of the satellite, which increases simultaneously the satellite visibility from both QGSs and the link attenuation, we included a MEO satellite at 8,000 km with a near-optimal inclination of 60 deg in the comparison. Table~\ref{tab3} shows the results obtained for these two orbits, and for 810 nm and 1550 nm with APD (Si-APD @ 810 nm or InGaAs-APD @ 1550 nm) and SNSPD detectors. For each set of parameters, we calculate the dual visibility time and communication time per day, and 
the averaged number of raw and distilled coincidences per day, all averaged over an observation time of 12 months. 
We assume clear sky, and day- and night-time operation at 1550 nm and only night-time operation at 810 nm, considering the stronger background noise at 810 nm. 
Night conditions are defined as both stations and the satellite not being illuminated by the Sun, also excluding dawn and twilight, as mentioned in subsection "Towards a resource demonstration system".
Our assumption of service availability during night time only at 810 nm explains the differences between the dual visibility time and the communication time per day for 810 nm in Table~\ref{tab3}.
We analyse the results in terms of performances only; cost should be taken into account in future iterations.

\begin{table*}[t]
\caption{
	Simulations results for a Low Earth Orbit  (600-km altitude) and a Medium Earth Orbit (8,000-km altitude). The LEO and MEO have an optimised inclinations of 50 and 60 degrees, respectively. 
	The averaging was carried out over the full 12 months of the simulation time span with time step equal to 10 s. 
	*) The dual link attenuations are averaged over the effective communication time during the simulation time span. Atmospheric losses are given by $\langle A \rangle$ (Eq.(\ref{Channel_Losses})) averaged over the communication time during the simulation time span and summed over the two channels. The system losses comprise the (principally absorption) losses in the ground- and space-based optical systems, as well as the photon detector efficiencies, and are given by $\eta_{sys}=(T_T \cdot T_R \cdot (1-L_P) \cdot T_{optics} \cdot PDE)^2$. The detectors are either SNSPDs, or APDs of the InGaAs (IGA) or Si type.
	}
	\begin{tabular*}{\linewidth}{lllllllll}
                  \hline
		 \hline
		 			   	                                           &   LEO 		 & LEO 	     		&  LEO	 &  LEO      &  MEO		    &  MEO	               &   MEO	 &   MEO		   \\
		 			   	                                           &   1550 nm       & 1550 nm 	     	&  810 nm	 &  810 nm  &  1550 nm     	    &  1550 nm	       &   810 nm	 &   810 nm           \\
		 			   	                                           &   SNSPD        & IGA-APD	 &  SNSPD	 &  Si-APD	  &  SNSPD	    &  IGA-APD    &   SNSPD	 &   Si-APD	   \\
		Average dual link attenuation (dB)*		   	 & 	66.0	          & 	77.2	     		&  63.8	 	 &  66.2	  & 	93.5	             & 104.7		               &    91.6		 &   94.0		           \\
		   of which: atmospheric losses $\langle A_1 \rangle + \langle A_2 \rangle$  (dB)		   	 & 	50.2	          & 	50.2     		& 41.9	 &  41.9	  & 	77.7	             & 77.7		               &    69.7		 &  69.7		           \\
		   of which: optical system losses $\eta_{sys}$ (dB)		 & 	15.8	          & 	27.0     		&  21.9	 &  24.3	  & 	15.8	          & 	27.0     		&  21.9	 &  24.3           \\
		Average dual visibility time per day (min)	 & 	9.3		 & 	 9.3	    		&  9.3		 &  9.3		  & 	153.1	     & 	 153.1              &   153.1	 &  153.1    	   \\
		Average communication time per day (min)	 & 	9.3		          & 	 9.3	    		&  2.3		 &  2.3  	  & 	153.1     	     & 	  153.1             &   4.1	 &   4.1  	   \\
		Average raw coincidences per day	   	         & 	8753	          & 	 676   		&  3858	  &  2203	  & 	237		     & 19 		       &  10	 	 &  6	   \\
		Average distilled coincidences per day	   	 & 	6038	          & 	 462  		&  2656	  &   1514	  & 	151		     & 	  9   		       &  6	 	 &  4	   \\
		\hline
		\hline
	\end{tabular*}
	\label{tab3}
\end{table*}

\begin{figure}[t!]
\begin{center}
\includegraphics[width=1 \columnwidth]{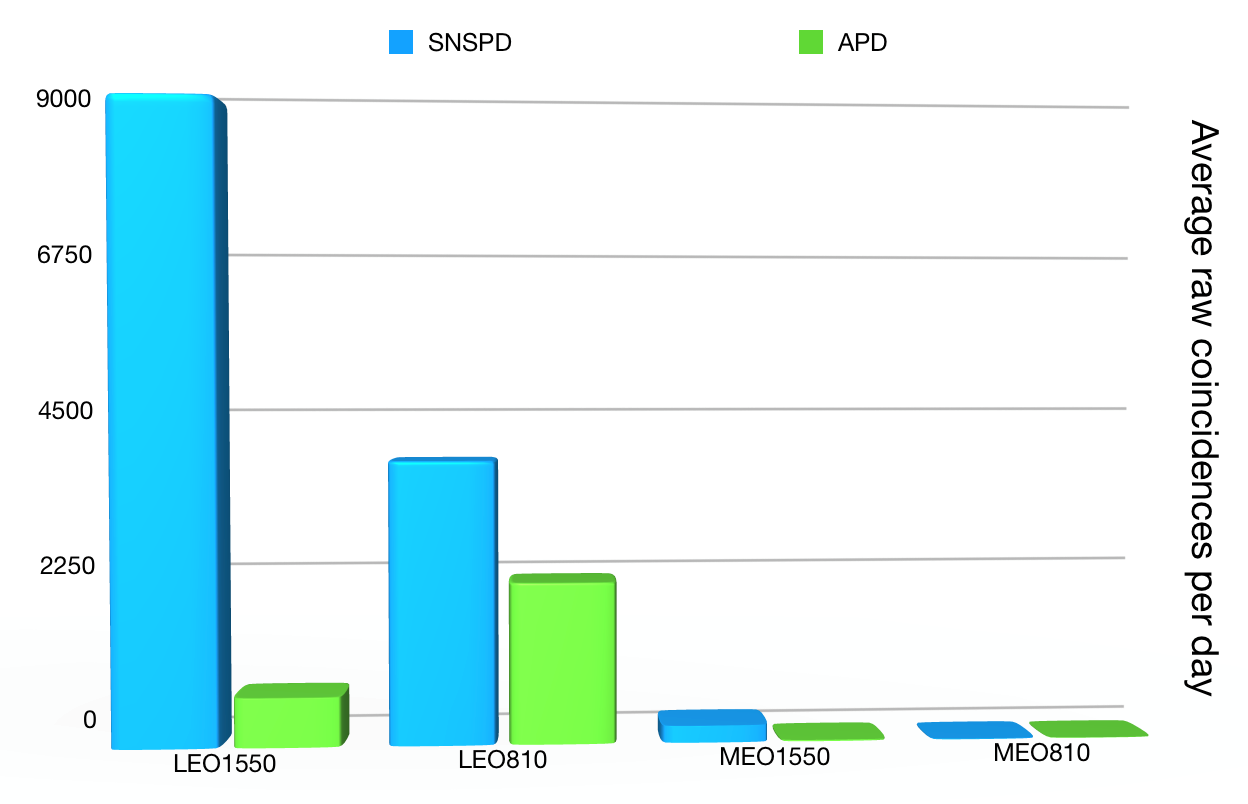}
\caption{Average raw coincidences per day as reported in Table~\ref{tab3}. Shown are results for a LEO with inclination 50 degree and 
height 600 km and a MEO with inclination 60 degrees and height 8,000 km, and for wavelengths of 810 nm and 1550 nm with APDs (Si-APD @ 810 nm, InGaAs-APD @ 1550 nm; green bars) and SNSPD (blue bars).}
\label{Figure6}
\end{center}
\end{figure}

Analyzing the results of Table~\ref{tab3}, we compare the average raw coincidences per day (Eq.(\ref{eq:coincidencesR})) as a function of the altitude in figure~\ref{Figure6}.

The number of raw coincidences is roughly 40 (400) times greater at 1550 nm (810 nm) for the LEO at 600 km than for the MEO at 8000 km. Our definition of night condition (both QGSs and the satellite in the dark) explains the large difference in the averaged communication time per day for the two wavelengths: the higher the satellite, the higher its exposition to the Sun. To summarise:

\noindent $\bullet$ Although MEO offers more than an order of magnitude longer dual visibility times, this advantage does not compensate for the much higher channel losses.

\noindent $\bullet$ SNSPD detectors enable more efficient operation than APD in all circumstances. In particular at 1550 nm InGaAs-APDs are to be avoided. 

\noindent $\bullet$ Despite the greater divergence of the light beam (diffraction) at 1550 nm, coincidences (raw, as well as distilled) are higher at 1550 nm than at 810 nm, provided SNSPD detectors can be used.

\noindent $\bullet$ If SNSPDs are not an option (e.g., because of complexity, form factor, cost, or low count rate), 810 nm in combination with Si-APDs is clearly the solution of choice.

\begin{figure}[t!]
\centering
\includegraphics[width=1 \columnwidth]{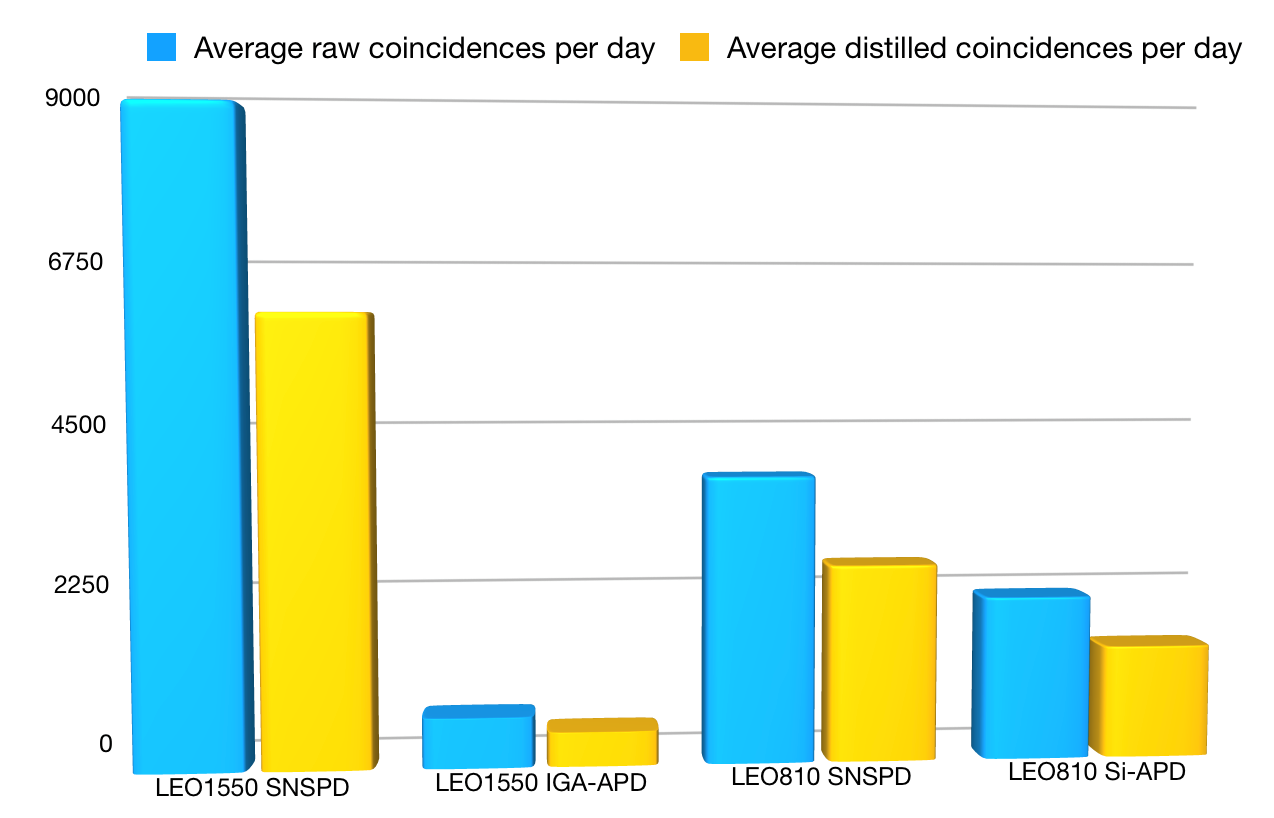}
\caption{Mean raw (blue bars) and distilled coincidences per day (orange) as reported in Table~\ref{tab3}. Shown are results for the 
LEO with inclination 50 degrees and height 600 km, and a wavelength of 810 nm and 1550 nm with Avalanche Photo Detectors (Si-APD @ 810 nm, InGaAs-APD @ 1550 nm) and Super-conducting Nanowire Single Photon Detectors.}
\label{Figure7}
\end{figure}

Focusing on the results for the LEO at 600 km, figure~\ref{Figure7} compares the rate of raw coincidences  (Eq.(\ref{eq:coincidencesR})), corrected for the basis reconciliation factor $q$, to the rate of distilled coincidences (Eq.(\ref{eq:distillation})). The difference between these two rates is due to rejection of false coincidences caused by dark- and background counts, multi-photon events generated in the spontaneous down-conversion process used to produce the entangled photon pairs, as well as imperfections allowing for the detection events to be registered by the wrong detector (polarisation errors), following the procedure outlined above \cite{Ma_2007}.

For the LEO, distilled coincidences are roughly equal to 69\% of the basis-reconciliation-corrected (i.e., multiplied by  $q=0.5$) raw coincidences, indicative of a good discrimination against dark- and background counts thanks to the relatively short coincidence time window (i.e., photon time tagging resolution), together yielding a relatively small QBER of about 2\%. The better choice remains a LEO at 600 km altitude at 1550 nm with SNSPD detectors, and in second position a LEO at 600 km altitude at 810 nm, also with SNSPD detectors. This conclusion should continue to hold,
even when taking into account the effects of cloud coverage that will affect both wavelengths observed here.  Much longer wavelengths would need to be considered in order to see a noticeable effect on cloud penetration, but the benefit would be countered by a strong increase of background photons. The potential of 1550 nm systems to operate under daytime conditions, on the other hand, strengthens the case for 1550 nm.

\begin{figure}[t!]
\centering
\includegraphics[width=1 \columnwidth]{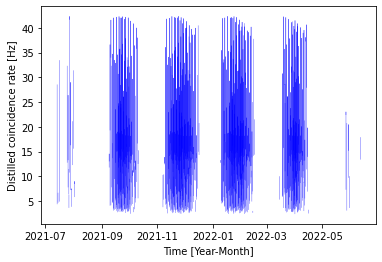}
\caption{Distilled coincidence rate for the Low Earth Orbit at 600 km with inclination of 50 degrees. The wavelength of 810 nm allows only night-time operation. The quantum ground stations are assumed to be equipped with Super-conducting Nanowire Single Photon Detectors. Simulation from July 1, 2021 to July 1, 2022 (meteorological effects not considered).}
\label{Figure8}
\end{figure}

\begin{figure}[t!]
\centering
\includegraphics[width=1 \columnwidth]{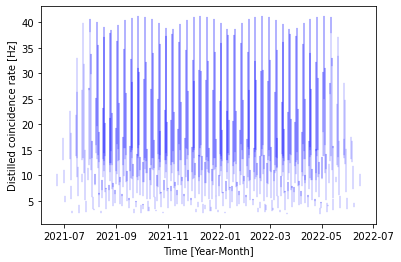}
\caption{Distilled coincidence rate for the Sun Synchronous Orbit at 600 km with inclination of 97.8 degrees. The wavelength of 810 nm allows only night-time operation. The quantum ground stations are assumed to be equipped with Super-conducting Nanowire Single Photon Detectors. Simulation from July 1, 2021 to July 1, 2022 (meteorological effects not considered).}
\label{Figure9}
\end{figure}

Finally, we demonstrate the importance of a long simulation duration of 12 months, in order to reveal seasonal and more subtle orbital effects, by plotting in figure~\ref{Figure8} the distilled coincidence rate for the simulation of the case of a LEO at 50 degrees inclination and a wavelength of 810 nm that allows, in our scenario, for night-time operation only. The effect of the short night duration in summertime is clearly visible, but even more striking is the complete absence of successful entanglement distribution during multiple relatively long periods of approximately one month each. The latter effect is due to the fact that the satellite or the QGSs are systematically illuminated by the Sun during multiple successive passages of the satellite over the QGSs. A simulation with an idealised satellite orbit, and even a realistic orbital simulation over a short number of days, might easily miss such important information. 
In fact, the behaviour of the LEO orbit shown in figure~\ref{Figure8}, could be a reason to consider an SSO, which shows a more regular distribution of successful events in time, as shown in figure~\ref{Figure9}. 
Of course, another reason to prefer an SSO despite its lower average performance (see Table \ref{tab4}) is the larger geographical coverage, extending to the highest latitudes.

\begin{table*}[t]
	\caption{
	Simulations results for a (frozen) Sun Synchronous LEO orbit at 600 km height with an inclination of 97.8 degrees.
	The averaging was carried out over the full 12 months of the simulation time span with time step equal to 10 s.
	*) The dual link attenuations are averaged over the effective communication time during the simulation time span. Atmospheric losses are given by $\langle A \rangle$ (Eq.(\ref{Channel_Losses})) averaged over the communication time during the simulation time span and summed over the two channels. The system losses comprise the (principally absorption) losses in the ground- and space-based optical systems, as well as the photon detector efficiencies, and are given by $\eta_{sys}=(T_T \cdot T_R \cdot (1-L_P) \cdot T_{optics} \cdot PDE)^2$. As before, the detectors are either SNSPDs, or APDs of the InGaAs (IGA) or Si type.
    }
	
  \centering
	\begin{tabular*}{0.7\linewidth}{lllll}
                  \hline
		 \hline
		 			   	                                           &   SSO 		 & SSO 	     		&  SSO	 &  SSO      		   \\
		 			   	                                           &   1550 nm       & 1550 nm 	     	&  810 nm	 &  810 nm            \\
		 			   	                                           &   SNSPD        & IGA-APD	     	&  SNSPD	 &  Si-APD	  	   \\
		Average dual link attenuation (dB)*	   & 	 66.2	 	           & 	77.3	  	                 &  63.8   	 &  66.3   	  	            \\
		   of which: atmospheric losses $\langle A_1 \rangle + \langle A_2 \rangle$  (dB)	   & 	 50.3	          & 	 50.3    		&    42.0  &   42.0     \\
		   of which:  optical system losses  $\eta_{sys}$ (dB)  & 	 15.8	          & 	27.0     		&  21.9	 &  24.3   \\
		Averaged dual visibility time per day (min)	          & 	2.97		 & 	 2.97    		&  2.97	 &  2.97	     	   \\
		Averaged communication time per day (min)	 & 	2.97	          & 	 2.97    		&  1.19	 &  1.19  	    	   \\
		Averaged raw coincidences per day	   	         & 	2796	          & 	 216	   		&   2052 	  &  1171  	     \\
		Averaged distilled coincidences per day	   	 & 	1929	          & 	 148  		&  1414	  &   806	     \\
		\hline
		\hline
	\end{tabular*}
	\label{tab4}
\end{table*}

The results of the simulations we have presented above give structuring orientations to the design of the demonstration system. Based on these orientations, we have gathered a preliminary list of requirements on the main subsystems of the demonstrators, that are ready to be shared with the entities that will be involved in their development. We also have identified the most critical subsystems in terms of impact on the overall performance and scheduling, and we have brought forward the development of two of these to make sure that they will be available in time.


\section{Conclusions}
\label{sec_Conclusions}

We have carried out the first steps of the design of a Quantum Information Network based on entanglement distribution from Space to ground. 

For this, we started by listing many possible use cases for industry, critical infrastructures, financial and administrative sectors, and operational and fundamental Science, as well as the main required key performance indicators for such a QIN. 

We propose a high-level architecture of a Space-based QIN composed of four main segments, the mission segment,  the control segment,  the Space segment, and the access segment. We explain how the entanglement resource is distributed across the  network towards final users, thus enabling the quantum-state teleportation between user end-nodes. The state-of-the-art of critical quantum subsystems of this architecture is discussed, particularly for entanglement photon sources, single photon detectors, and quantum memories. The current status of standardisation frameworks is also presented.

Based on these elements, we expose an explicit and effective 15-year roadmap to create in three stages an operational QIN service. The first stage demonstrates entanglement resource distribution from Space, while academic actors mature key devices to handle the resource on the ground. The second stage demonstrates the construction of a long distance experimental network that assembles several links that use  Space-distributed entanglement. The third stage consists of building an operational system by using the outcomes of the first two stages. 

Finally, we describe how we have already begun executing this roadmap by designing a demonstration system for the resource stage, carrying out simulations in order to make a number of system trade-offs on design drivers and defining a first level of requirements on the main, critical subsystems of this demonstrator. 

The next step of this work is to continue with the resource stage of our roadmap, refining the design of the demonstration system up to detailed plans, and from there to develop it further by
considering a constellation of satellites \cite{Khatri_2021}. 
We have anticipated work on the most critical elements, and will be happy to collaborate with entities that have the expertise required to contribute to the various elements of this system. We count on research teams to improve, in parallel, the core technologies of entanglement storage and swapping at the network stage.

\section{Methods}
\label{sec_Methods}

The atmospheric link simulation code first applies the radar equation to each single downlink between the satellite and the QGSs \cite{Kerstel_2018, Pfennigbauer_2005}.  
For a single link $x$ ($x=1,2$), the instantaneous atmospheric link attenuation (i.e., the inverse of the atmospheric transmittance) at time $t$ between the 
satellite optical terminal's exit and the QGS's telescope entrance (i.e., its first optical surface) is given by:

\begin{eqnarray}
  A_x(t) = \frac{L_x(t)^2(\theta_{diff}^2+\theta_{atm}^2)}{D_R^2} 10^{A_{atm,x}(t)/10}~,
\label{Channel_Losses}
\end{eqnarray}

\noindent with $L$ the link distance between the QGS and the satellite, $D_R$  the receiver telescope diameter, 
and $A_{atm,x}$ the atmospheric attenuation due to scattering  and  absorption, expressed in dB. $A_{atm,x}$ is a function of the elevation angle $\beta(t)$: $A_{atm,x}(t) = A_{atm, 0} / \sin{\beta_x(t)}$, with $A_{atm,0}$ the atmospheric attenuation at zenith. We assume $A_{atm,0} =3$ dB at 810 nm and  $A_{atm,0} =2$ dB at 1550 nm. 

The angles $\theta_{diff}=2.44 \lambda/D_T$ and $\theta_{atm} = 2.1 \lambda/r_0$ are, respectively, the diffraction limited and atmospheric turbulence induced divergence angles of the satellite's transmitter telescope, with $\lambda$ the quantum signal wavelength, $D_T$  the transmitter telescope diameter, and $r_0$ the Fried parameter. Atmospheric turbulence effects are considered to be negligible in the downlink configuration. Consequently, we set $r_0  \gg D_R$. 

The parameters used for the calculation of the link budget are summarised in Table~\ref{tab2}.

As we will see, with our assumptions for a satellite in LEO, the purely atmospheric attenuation is about 8.3 dB higher at 1550 nm than at 810 nm.
This difference is mainly due to the higher beam divergence at 1550 nm.

Subsequently, the overall quantum channel transmittance efficiencies can be written:

\begin{eqnarray}
   \eta_x = T_T \cdot T_R \cdot (1-L_P) \cdot T_{optics} \cdot PDE / A_x(t)
\label{eq:OverallChannelEfficiency}
\end{eqnarray}

These are inserted in the model developed by Ma \textit{et al}. (2007) \cite{Ma_2007} 
that gives the probability of a coincidence detection in channel $x=1, 2$. 
Here we have introduced the pointing losses through the parameter $L_P$, as well as the losses in the optical systems through the parameters $T_T, T_R$, and $T_{optics}$. The pointing losses are estimated to be equal to 0.3 at 810 nm and 0.2 at 1550 nm (corresponding to a pointing precision of $\sim10~\mu$rad and $\leq10~\mu$rad, respectively). The parameters $T_T$ and $T_R$ model the optical transmission of the sending and receiving telescopes, respectively, whereas $T_{optics}$ models the transmission of all the optics between the telescope and the detectors (i.e., essentially the detection modules in the QGSs and the optics between the entangled pair source and the satellite's telescopes).
For the estimation of $T_R$, $T_T$, and $T_{optics}$, we assume single-mode fibre-coupling at both receiver and transmitter modules. 
The values reported in Table~\ref{tab2} are likely to be rather conservative. 
This has the advantage that we do not necessarily need to assume that the single-mode fibre-coupling at the receiver is complemented with adaptive optics to improve the coupling efficiency (although a tip/tilt will still be required). 
Fibre coupling is practically mandatory when using SNSPD detectors on the ground, while fibre-coupling may not be a must when using APDs. 
Using APDs without fibre-coupling could lead to better performances, particularly at 810 nm with Si-APDs. 
We note that single-mode fibre coupling in satellite-to-ground laser links is a tedious task, due to the required correction of aberrations in real time, that has been modelled in Refs.~\cite{Canuet_2018, Scriminich_2021, Valentina_Marulanda_Acosta_2021}, and implemented in Ref.~\cite{Giggenbach_2022}. 

While this model has been developed for QKD using a pulsed entangled photon source, its results are practically indistinguishable from those of the recently published model for a continuous entangled photon source, as long as the pair production rate is not very large ($\lessapprox$ 100 MHz) \cite{Neumann_2021, Scheidl_2009}.

The model gives the probability of a coincidence detection between QGS$_1$ and QGS$_2$ as:

\begin{eqnarray}
\nonumber
  Q(\mu)= 1-\frac{1-Y_{01}}{(1+\eta_1 \frac{\mu}{2})^2} - \frac{1-Y_{02}}{(1+\eta_2 \frac{\mu}{2})^2}\\
  +\frac{(1-Y_{01})(1-Y_{02})}{(1+(\eta_1+\eta_2 - \eta_1  \eta_2 ) \frac{\mu}{2})^2} .
\label{Proba_detection}
\end{eqnarray}

Here $\mu$ is the average number of photon pairs per time window of interest (e.g., pulse), and $\eta_{1,2}$ are the efficiencies of Eq.(\ref{eq:OverallChannelEfficiency}). Furthermore, $Y_{0x}$ represents the detector noise for channel $x = 1,2$, such that $Y_{0x}=(4D+B)\tau$, with $D$ the detector dark count rate, $B$ the background (stray light) count rate, and $\tau$ the coincidence time window (we assume 4 detectors per station). With these definitions the rate of (raw) coincidence detection becomes:

\begin{equation} \label{eq:coincidencesR}
R_{coinc}=\frac{Q(\mu)}{\tau}~,
\end{equation}
whereas the pair production rate is given by $R_{pair}=\mu/\tau$, and is generally determined by the entangled photon source characteristics. The raw coincidence rate given by Eq.(\ref{eq:coincidencesR}) includes false coincidences due to dark- and background counts, as well as imperfections allowing for the detection events to be registered by the wrong detector (e.g., polarisation errors). We estimate here the rate of the distilled coincidences, i.e., the useful coincidences after false coincidence filtering, by analogy with post-processing for quantum keys distribution, as in, e.g., Ref.~\cite{Ma_2007}. According to this approach, a lower limit for the efficiency with which the number of useful coincidences that can be obtained after post-processing is given by:
\begin{eqnarray}
\nonumber
R_{dist}(QBER) \geq q ( 1- f(QBER) H_2(QBER)\\
-H_2(QBER) )
\label{eq:distillation}
\end{eqnarray}

This model does not take into account finite-size effects of the key length, which will reduce this lower bound in practical system implementation. The basis reconciliation factor $q$ in Eq.(\ref{eq:distillation}) describes the random selection of one of the two polarisation bases by the two QGSs. We will assume $q=0.5$, as for the BBM92 protocol \cite{BBM92protocol}. The bidirectional error correction efficiency $f(QBER)$ is in general a function of the Quantum Bit Error Rate (QBER). It equals unity in the Shannon limit. Here we chose a conservative value of $f(QBER)=f=1.22$, following \cite{Ma_2007}. 
The binary entropy function is defined as: $H_2(x)=-x  \log_2(x) -(1-x)  \log_2(1-x)$. Finally, calculation of the distillation efficiency requires knowledge of the QBER, which can be determined experimentally. Instead, we use the estimation \cite{Ma_2007}:

\begin{eqnarray}
QBER =e_0- \frac{C}{Q(\mu)} ~,
\label{eq:QBER}
\end{eqnarray}
with 

\begin{eqnarray}
C =  \frac{(e_0-e_p)\eta_1\eta_2\mu \left( 1+\frac{\mu}{2} \right)}{(1+\eta_1 \frac{\mu}{2})(1+\eta_2 \frac{\mu}{2})(1+ ( \eta_1 +\eta_2 -\eta_1 \eta_2) \frac{\mu}{2})}~,\ \
\label{eq:QBERnumerator}
\end{eqnarray}
where $e_0$ is the error probability for dark- and background counts, whereas $e_p$ is the polarisation error for entangled photon detection. 
A full list of parameters used in our simulations is provided in Table~\ref{tab2}.
\newline
\newline

\noindent \textbf{Acknowledgements}

\noindent This work was supported by the French Space Agency, the Centre National d'Etudes Spatiales (CNES), and Thales Alenia Space France. It relies on numerous prior works supported by the European Commission, the French Ministery of Defence (DGA), and the French National Research Agency (ANR). The authors thank Julien Laurat for useful discussions on quantum memories and Luca Lazzarini for his assistance with double-checking the numerical simulation results.
\newline

\noindent \textbf{Data availability}

\noindent All data presented in this paper can be reproduced by following the described methodology. 
\newline

\noindent \textbf{Code availability}

\noindent The code used to generate the demonstrator simulation data is available upon reasonable request, to be addressed to E.K. The Simu-CIC code for satellite orbital propagation is available from CNES (see the text for details).
\newline

\noindent \textbf{Competing interests}

\noindent The authors declare no competing interests. E.K. is an Editor of the Communications Physics collection on Space Quantum Communications, but was not involved in the editorial review of, or the decision to publish this article.
\newline

\noindent \textbf{Author contributions}

\noindent The study was conceived and designed by LFP, ED, PG, EK, ST, and MvdB. OA, AL, AMa, and TT principally performed the subsystem hardware analyses led by ST, whereas ED and MS principally studied protocol issues. JD, SG, and AMe performed the orbital link efficiency studies coordinated by EK and LFP. LFP, PG, and MvdB focused on the overall system architecture. LFP and MvdB coordinated the overall study. The first draft of the manuscript was written by LFP, EK, and MvdB. All authors read and approved the final manuscript.
\newline

\end{document}